\documentclass[10pt,journal,compsoc]{IEEEtran}
  \usepackage[nocompress]{cite}
  \usepackage{cite}
  \usepackage{amsmath,amsfonts}
  \usepackage{algorithmic}
  \usepackage{algorithm}
  \usepackage{array}
  \usepackage{color}
  \usepackage[caption=false,font=normalsize,labelfont=sf,textfont=sf]{subfig}
  \usepackage{textcomp}
  \usepackage{stfloats}
  \usepackage{url}
  \usepackage{verbatim}
  \usepackage{graphicx}
  \usepackage{amssymb}
  \usepackage{bm}
  \usepackage{cite}
  \usepackage{booktabs}
  \usepackage{tabularx}
\ifCLASSINFOpdf

\else

\fi
\begin{document}

\title{Near-Field Directional Modulation for RIS-Aided Movable Antenna MIMO Systems with Hardware Impairments}

\author{Maolin Li, Feng Shu, Riqing Chen, Cunhua Pan, Yongpeng Wu
	\thanks{This work was supported in part by the National Natural Science Foundation of China under Grant U22A2002, and  by the Hainan Province Science and Technology Special Fund under Grant ZDYF2024GXJS292; in part by the Scientific Research Fund Project of Hainan University under Grant KYQD(ZR)-21008; in part by the Collaborative Innovation Center of Information Technology, Hainan University, under Grant XTCX2022XXC07; in part by the National Key Research and Development Program of China under Grant 2023YFF0612900. (Corresponding author: Feng Shu.) }
	\thanks{Maolin Li is with the School
		of Information and Communication Engineering, Hainan University, Haikou,
		570228, China. (e-mail: limaolin0302@163.com).}
	\thanks{Feng Shu is with the School of Information and Communication Engineering and Collaborative Innovation Center of Information Technology, Hainan University, Haikou 570228, China, and also with the School of Electronic and Optical Engineering, Nanjing University of Science and Technology, Nanjing 210094, China. (e-mail: shufeng0101@163.com).}
	\thanks{Yongpeng Wu is with the Shanghai Key Laboratory of Navigation and
		Location Based Services, Shanghai Jiao Tong University, Minhang, Shanghai,
		200240, China. (e-mail: yongpeng.wu2016@gmail.com).}
	\thanks{Riqing Chen is with the Digital Fujian Institute of Agricultural Big Data, Fujian Agriculture and Forestry University, Fujian, 350002, China, and also with the Fujian Key Lab of Agricultural IOT Applications, Sanming University, Fujian, 365004, China. (e-mail: riqing.chen@fjsmu.edu.cn).}
	\thanks{Cunhua Pan is with the National Mobile Communications Research Laboratory,
		Southeast University, China. (e-mail: cpan@seu.edu.cn).}
}

\IEEEtitleabstractindextext{%
\begin{abstract}
Movable antennas (MAs) are a promising technology to achieve a significant enhancement in rate for future wireless networks. The pioneering investigation on near-field directional modulation design for a reconfigurable intelligent surface (RIS)-assisted MA system is presented, with the base station equipped with a MA array. To maximize the secrecy sum rate (Max-SSR) with hardware impairments (HWIs) and imperfect channel state information (CSI), which involves a joint optimization of beamforming vectors for confidential messages and artificial noise (AN), power allocation factors, phase shift matrices, MA positions, and receive beamforming vectors. Firstly, the transmit beamforming vectors and phase shift matrices are iteratively optimized, leveraging leakage theory and phase alignment techniques. Then, two novel algorithms for discrete MA positioning are proposed, respectively, employing uniform and compressed sensing (CS)-based non-uniform grouping strategies. Subsequently, the AN is considered and designed as the additional energy required for zero-space projection, and the receive beamforming vector is derived using the minimum mean square error (MMSE) method. The proposed algorithms have low computational complexity. Simulation results demonstrate the effectiveness of the proposed algorithms. Under HWIs and imperfect CSI, the proposed algorithm can achieve a 28\% enhancement in SSR performance while reducing the number of antennas by 37.5\% compared to traditional fixed-position antenna (FPA) systems.
\end{abstract}

}

\maketitle

\IEEEdisplaynontitleabstractindextext

\IEEEpeerreviewmaketitle

	\section{Introduction}
\IEEEPARstart{W}{ith} the advancement of 6G communication systems, strengthening information security has become imperative. Mathematical encryption at the architectural level serves as a critical technology for enhancing system security. However, conventional cryptographic methods incur substantial computational overhead and key management complexity to counter eavesdroppers (Eve) with evolving computing capabilities. Driven by the dual demands of cost efficiency and security enhancement, physical layer security (PLS) technologies have garnered significant attention as a promising paradigm. PLS techniques employ methodologies including beamforming~\cite{Chen2019}, artificial noise (AN)~\cite{Shu2021a}, and channel state information (CSI) exploitation to enhance confidentiality through intrinsic wireless channel properties~\cite{Lin2024}. In particular, the rapid development of multi-input multi-output (MIMO) systems has promoted applying PLS technologies in analogue, digital, and hybrid beamforming structures, and extensively studied the design methods of antenna weights for maximizing secrecy rate (SR). These studies have demonstrated that beamforming/precoding techniques with well-designed antenna weights can improve system security.

\subsection{Background}

Directional modulation (DM), as a PLS technique rooted in Shannon's information-theoretic foundations~\cite{Shu2024a}, orchestrates secure symbol transmission through multidimensional control in time~\cite{ManeiroCatoira2024}, spatial~\cite{Shu2023}, and frequency domains~\cite{Shu2018}. By exploiting inherent channel discrepancies between legitimate users (Bob) and Eve, DM ensures correct demodulation at authorized Bob's receivers while distorting signal constellations at Eve terminals. This paradigm can also be achieved by utilizing additional power, for example, the concept of AN was first proposed in~\cite{Goel2008} and applied to DM design. In~\cite{Wei2021}, three optimized symbol modulation schemes were developed using the minimum Euclidean distance criteria and priority-aware scheduling, effectively degrading Eve's interception capabilities. While existing studies have illustrated superior performance in point-to-point communication scenarios, security vulnerabilities emerge when the Eve channel gain exceeds that of the Bob~\cite{Qiu2024}, particularly in situations where Eve is positioned closer to the base station (BS). In such cases, the reduced path loss enables the eavesdropper to establish a more favourable transmission link. To solve this issue, an additional scope of security assurance was created using a distributed DM architecture in~\cite{Qiu2024}, but its feasibility in the communication near-field (NF) has not been studied. Moreover, the power consumption of the distributed DM structure is very high, making it difficult to be applied in practice.

Due to the limitations of traditional schemes in improving security performance and reducing costs, reconfigurable intelligent surfaces (RISs) have garnered extensive attention~\cite{Shi2023}. Low-cost RISs boost communication performance and security by enabling additional transmission paths and offering the flexibility to reconfigure wireless channels~\cite{Ju2025}. For example, in~\cite{Cheng2024}, a multi-access scheme with the highest degrees of freedom (DoF) was designed by exploiting RIS phase modulation. The analysis shows that RIS can significantly increase the DoF of the system in both the presence and absence of direct links. To address the trade-off between hardware cost and security performance, a method based on semi-positive definite relaxation, alternating direction multiplier, and user scheduling strategy was proposed in~\cite{Li2024}. By jointly designing active beamforming and RIS, the mixed integer problem caused by low-resolution digital-to-analogue converters was solved, and the SR was maximized. Conventional passive RIS are fundamentally limited to phase optimization of signals, whereas active RIS exhibits significantly higher power consumption and implementation complexity~\cite{Le2025}. To achieve trade-offs between hardware efficiency, energy constraints, and DoF, the innovative absorption RIS was proposed in~\cite{Wang2024a} to enhance PLS. 
To expand the coverage, a secure communication system with simultaneous transmission and reconfigurable intelligent surface (STAR-RIS) assistance was studied in~\cite{Zhu2025}, and two reinforcement learning algorithms were proposed to improve the long-term average SR. The deployment of RIS reduces the cost of traditional DM systems while delivering substantial security performance gains in practice.

As previously mentioned, enhancing spatial freedom can improve security performance. Recently, movable antennas (MAs) have garnered significant attention due to their ability to move within a specified range, offering new degrees of freedom (DoF) to fully leverage the spatial variations of wireless channels. The MAs technology originated from the micro-electromechanical systems (MEMS)-based reconfigurable antenna proposed in \cite{Chiao1999}, which dynamically modulates radiation patterns through precise adjustment of the vee antenna. In \cite{Kosta2004}, a liquid antenna system using mercury as the conductive liquid was introduced. Subsequently, the concepts of MA and fluid antenna (FA) were proposed in \cite{2008} and \cite{Wong2020}, respectively. Following this, MA-assisted communication was presented in \cite{Zhu2024b} and demonstrated to have better performance compared to fixed position antenna (FPA). Driven by this, it is an interesting issue to conduct DM design by taking advantage of the DoF brought by the MA system, thereby reducing costs and improving safety.

\subsection{Motivations and Contributions}

In summary, although DM and RIS technologies have been widely studied, the security performance of MA-assisted NF communication is still in its infancy. To the best of our knowledge, DM for MA systems has not yet been proposed, and the existing discrete MA location optimization methods have excessively high computational complexity. To address these challenges, this paper pioneering investigates the DM design of RIS-assisted communication systems with MA in NF.
The contributions of this paper are summarized as follows:
\begin{itemize}
	
	\item We study the NF DM design for RIS-assisted MA systems under HWIs and imperfect CSI. Considering discrete MA positions and a limited range of movement, a secrecy sum rate (SSR) maximization problem is formulated subject to constraints on transmit power, receive power, artificial noise, MA positions and quantities, and RIS reflective power.
	
	\item To solve the non-convex optimization problem, we decompose the original problem into tractable subproblems and develop three algorithms. Firstly, an iterative method based on leakage theory and phase alignment is proposed to optimize the transmit beamforming vectors and phase shift matrix, which address the phase mismatch issues in multi-user scenarios. Regarding the discrete MA positioning, we propose uniform and non-uniform grouping algorithms that significantly reduce computational complexity compared to exhaustive search methods, respectively.
	
	\item Simulation results demonstrate that under conditions of HWIs and imperfect CSI, the MA array achieves higher SSR gains compared to traditional FPA systems, while requiring fewer antenna elements. Furthermore, it is observed that active RIS provide significantly greater SSR gains than passive RIS counterparts. In MA systems, SSR performance initially improves with an increase in the number of antennas, candidate positions, and transmit power, before eventually reaching a saturation point. Additionally, the performance enhancement from increasing the number of RIS elements yields relatively low SSR gains in MA-aided systems.
\end{itemize}	
\subsection{Organization and Notation}

The remaining part of the paper is as follows. Sec. \ref{sec:2} reviews the related work. In Sec. \ref{sec:3},  the system model and problem formulation are presented. In Sec. \ref{sec:4}, three algorithms are proposed to solve the non-convex problem. Simulation results are provided in Sec. \ref{sec:5}, followed by the conclusion in Sec. \ref{sec:6}.

Notations: a, $\mathbf{a}$, and $\mathbf{A}$ represent a scalar, a vector, and a matrix, respectively. $\mathcal{A}$ represents a set. $\mathbb{C}^{M\times N}$ denotes the space of $M\times N$ complex matrix. $|\ |$, $\|\ \|_0$, $\|\ \|_1$, $\|\ \|$, and $\|\ \|_{F}$ denote the modulus operation, $l_0$-norm, $l_1$-norm, $l_2$-norm, and Frobenius norm, respectively. $[\ ]^T$, $[\ ]^*$, and $[\ ]^H$ are the transpose, conjugate, and conjugate transpose, respectively. $\cdot$ and $\otimes$ represent dot product and Kronecker product, respectively. $\mathbb{E}\{\ \}$ stands for expectation. $\text{diag}(\ )$ denotes the diagonal operator. $[\ ]^+$ is a value greater than or equal to zero.
\section{Related Works}\label{sec:2}	
In this section, we comprehensively survey related work on MA technologies. Specifically, we examine existing MA positioning optimization methods and security-driven MA techniques, with comparative analysis highlighting our novel contributions.

MAs offer significant benefits, including enhanced communication~\cite{Zhu2024d}, flexible beamforming~\cite{Yang2024,Zhu2023}, and spatial multiplexing~\cite{Sun2025}, which provide innovative solutions for adaptive communication environments. These capabilities enable superior performance in complex scenarios, driving advancements in mobile communication technology and facilitating the integration of multiple technologies in 6G networks. In~\cite{Tang2025a, Hu2024a}, by jointly optimizing the MA's positions, AN and transmit beamforming, an element-by-element method was proposed to solve the SR maximization problem, and it was proved that MA is superior to FPA in terms of security. An improved particle swarm optimization algorithm was proposed in~\cite{Ding2025} to optimize the MAs' positions and maximize the sum of SR to counter multiple Eves. A block coordinate descent algorithm was proposed in~\cite{Liu2025, Mao2025} to solve the maximum SR problem with the minimum spacing constraint of MAs. In~\cite{Hu2024b}, the secrecy outage probability in a practical scenario without Eve's channel state information (CSI) was studied, and the projection gradient descent algorithm was used to optimize the positions of MAs~\cite{Hu2025}. For MA-assisted communication systems with RIS~\cite{Zhang2025}, a novel deep reinforcement learning (DRL) method was proposed in~\cite{Xie2025} to optimize the motion trajectories of MAs, achieving a higher covert rate than FPA. In~\cite{Kang2024}, a deep learning model was proposed to optimize antenna positions and beamforming. Optimizing the positions of MAs is crucial in MA-assisted communication systems. 

 The aforementioned studies primarily consider scenarios with continuously adjustable MA positions. In practical deployments, MAs are constrained to discrete positioning configurations. To address this implementation constraint, existing studies have investigated MA system performance under discrete position constraints. For example, the six-dimensional movable antenna (6DMA) was proposed in~\cite{Shao2025,Shao2025a}, which can rotate and adjust positions in 3D space, offering more DoF than traditional FPA and FA. Additionally, an iterative optimization algorithm was proposed to solve the 3D position/3D rotation of each 6DMA surface, significantly enhancing transmission rates. References~\cite{Wu2023} and~\cite{Mei2024} employed distinct methodologies for optimizing discrete MA positions to enhance secure transmission performance, with~\cite{Wu2023} utilizing an iterative approach based on generalized Bender's decomposition and~\cite{Mei2024} developing a graph-based optimization technique. However, existing approaches exhibit prohibitively high computational complexity.

Related works with continuous or discrete MAs have demonstrated that MAs consistently outperform FPAs in security performance. However, several practical constraints must be addressed. In next-generation communications, as the number of antennas increases and antenna arrays become larger in physical size or are deployed at higher frequencies, the NF boundary is significantly extended~\cite{Yadav2024}. Therefore, NF communications are expected to be more widespread in the future. To the best of our knowledge, for MA-assisted communication scenarios in NF, the problems of maximizing transmission rate and minimizing power consumption have been investigated in~\cite{Ding2025a} and~\cite{Zhu2025a}, respectively. However, security performance has not been examined. Research on the security performance of MA-assisted systems in the NF remains at a nascent stage. The discrete movement constraints of MAs necessitate the development of low-complexity position optimization algorithms to achieve an optimal trade-off between computational complexity and security enhancement. Furthermore, hardware impairments (HWIs) and imperfect CSI are issues that need to be considered in practical deployment~\cite{Bai2024,Zhai2024,Wang2024c}. In Table \ref{tab}, our main concerns and comparisons with related work are summarized.
\begin{table*}[t]
	\centering
	\caption{Our main focuses and comparisons with related works.}\label{tab}
	\renewcommand{\arraystretch}{1.2}
	\setlength{\tabcolsep}{8pt}
	\begin{tabularx}{\linewidth}{c|c|c|c|c|c|c|c}
		\toprule
		Related works & RIS & Near field & Security &HWIs & Discrete & Method                   & Complexity \\		\midrule
		\cite{Tang2025a,Liu2025,Mao2025}      &$\times$ &$\times$&\checkmark&$\times$&$\times$& Block coordinate descent & Moderate     \\\hline
		\cite{Hu2024a,Hu2024b,Hu2025}      & $\times$ &$\times$&\checkmark &$\times$&$\times$          & \begin{tabular}[c]{@{}c@{}}Projected gradient ascent and\\  the alternating optimization methods\end{tabular}  &Moderate            \\\hline
		\cite{Ding2025}     & $\times$ &$\times$&\checkmark&$\times$&$\times$          &Multi-velocity particle swarm optimization&Higher   \\     \hline  
		\cite{Zhang2025}     & \checkmark &$\times$&$\times$&$\times$&$\times$          &Genetic algorithm&Higher   \\     \hline 
		\cite{Xie2025}     & \checkmark &$\times$&\checkmark&$\times$&$\times$          &Deep reinforcement learning&Higher   \\     \hline 
		\cite{Shao2025a}     &$\times$ &$\times$&$\times$&$\times$&\checkmark& Block coordinate descent & High     \\\hline   
		\cite{Wu2023}      & $\times$ &$\times$&$\times$&$\times$&\checkmark          &Generalized Bender's decomposition&High   \\     \hline 
		\cite{Mei2024}      & $\times$ &$\times$&$\times$&$\times$&\checkmark          &A graph-based approach&Higher   \\     \hline  
		\cite{Ding2025a}      & $\times$ &\checkmark&$\times$&$\times$&$\times$          &\begin{tabular}[c]{@{}c@{}}Dynamic neighborhood pruning\\  particle swarm optimization\end{tabular}  &Higher   \\     \hline 
		\cite{Zhu2025a}      & $\times$ &\checkmark&$\times$&$\times$&$\times$          &MA design strategies based on statistical CSI  &lower   \\     \hline    
		Ours      & \checkmark &\checkmark&\checkmark&\checkmark&\checkmark          &Algorithm \ref{alg:alg2} and Algorithm \ref{alg:alg3}&lower   \\     \hline    
	\end{tabularx}
\end{table*}
\section{System Model and Problem Formulation}\label{sec:3}
\subsection{System Model}
\begin{figure}
	\centering
	\includegraphics[width=0.45\textwidth, trim = 2 10 2 2,clip]{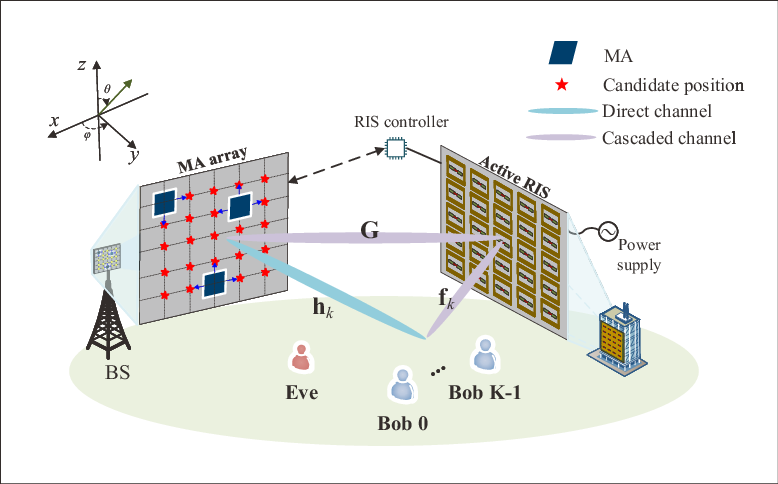}\\
	\caption{RIS-aided MA communication model.}\label{fig:1}
\end{figure}
The architecture of the RIS-aided multi-user MIMO downlink communication MA system is shown in Fig. \ref{fig:1}. The BS is equipped with $N_a$ MAs connected to the RF chains via flexible wires, and an active RIS with $M=M_hM_v$ reflecting elements is deployed, where $M_h$ and $M_v$ represent the number of horizontal and vertical RIS units, respectively. The BS transmits confidential messages (CMs) to $K$ Bob via both direct and cascaded channels, with each Bob equipped with $N_k$ antennas. Here, the cascaded channel denotes the links reflected by RIS. The positions of MAs with a minimum spacing of $d_{\min}$ are adjusted within the movable range (MR) $\mathcal{C}_{B}=N_hd_{\min}\times N_vd_{\min}$, where $N_h$ and $N_v$ denote the number of horizontal and vertical candidate positions, respectively. Since the positions of the MAs are discrete, the number of candidate positions can be expressed as $N=N_hN_v$. A three-dimensional (3D) coordinate system is considered. The origin is placed at the centre of $\mathcal{C}_{B}$. MAs, RIS, and Bob are located at the x-o-z, y-o-z, and x-o-y planes, respectively. Eve, equipped with $N_e$ antennas, is considered to be located near Bob. Given the development trend of large-scale antennas and the deployment of RIS near Bob, we assume that both Bob and Eve are located in the NF. For notational simplicity, we define sets $\mathcal N_a = \{0,1,...,N_a-1\}$, $\mathcal K = \{0,1,...,K-1\}$, $\mathcal N_h = \{0,1,...,N_h-1\}$, $\mathcal N_v = \{0,1,...,N_v-1\}$, $\mathcal M_h = \{0,1,...,M_h-1\}$, $\mathcal M_v = \{0,1,...,M_v-1\}$, $\mathcal M = \{0,1,...,M-1\}$, and $\mathcal N = \{0,1,...,N-1\}$ to collect the indices of the MAs, Bob, candidate positions on the x-axis, candidate positions on the z-axis, RIS units on the y-axis, RIS units on the z-axis, RIS units, and candidate positions.

The transmitted signal with HWIs can be represented as~\cite{Bai2024} 
\begin{equation}
	\label{eq1}
	\mathbf{x}=\sum_{k=0}^{K-1}(\sqrt{\tfrac{\alpha P_0}{K}}\mathbf{v}_ks_k+\sqrt{\tfrac{(1-\alpha) P_0}{K}}\mathbf{v}_{a,k}z)+\mathbf{z}_t,
\end{equation}
where $P_0$ denotes the transmission power. $\alpha$ denotes the power allocation factor for CM $s_k$ with $\mathbb{E}[|s_k|^2]=1$ and AN $z$ with $\mathbb{E}[|z|^2]=1$, satisfying $0\leq\alpha\leq1$. $\mathbf{v}_k\in\mathbb{C}^{N\times1}$ and $\mathbf{v}_{a,k}\in\mathbb{C}^{N\times1}$ are the beamforming vectors of $s_k$ and $z$, $\mathbf{v}_{k}^H\mathbf{v}_{k}=1$ and $\mathbf{v}_{a,k}^H\mathbf{v}_{a,k}=1$, respectively. $ \mathbf{z}_{t}$ stands for the distortion noise of Gaussian distribution, which is proportional to the average power of the transmission, i.e., $ \mathbf{z}_{t} \sim \mathcal{CN}(0,\mu_{t}\mathbf{R}_t)$, where $\mu_{t}$ is the ratio of distortion noise power to transmission power and
\begin{equation}
	\label{eq2}
	\mathbf{R}_t=\widetilde{\text{diag}}(\sum_{k=0}^{K-1}{\tfrac{\alpha P_0}{K}}\mathbf{v}_k\mathbf{v}_k^H+{\tfrac{(1-\alpha) P_0}{K}}\mathbf{v}_{a,k}\mathbf{v}_{a,k}^H). 
\end{equation}

The channels from BS to RIS, from RIS to Bob $k$, and from BS to Bob are denoted as $\mathbf{G}\in\mathbb{C}^{M\times N}$, $\mathbf{F}_k\in\mathbb{C}^{M\times N_k}$, and $\mathbf{H}_k\in\mathbb{C}^{N\times N_k}$, respectively. Similarly, $\mathbf{F}_e$ and $\mathbf{H}_e$ represent the channels from RIS to Eve and from BS to Eve, respectively. Then, the signal received at Bob $k$ can be expressed as
\begin{align}
	\begin{split}
		\label{eq3}
		\mathbf{y}_k&=(\mathbf{H}_k^H+\mathbf{F}_k^H\boldsymbol{\Theta}\mathbf{G})\mathbf{T}\mathbf{x}+\mathbf{n}_k+\mathbf{z}_{r,k}+\mathbf{F}_k^H\boldsymbol{\Theta}\mathbf{n}_r,
	\end{split}
\end{align}
where $\boldsymbol{\Theta}=\text{diag}(\bm{\theta})$ denotes the phase shift matrix. $\bm{\theta}=[\varrho_{_0} e^{j\varpi_0}, \varrho_{_1} e^{j\varpi_1},\ldots,\varrho_{_{M-1}} e^{j\varpi_{M-1}}]^T\in\mathbb{C}^{M\times 1}$, where $\varrho_m>1$ and $\varpi_m\in[0\ 2\pi]$ are adjustable amplitude and phase of RIS, respectively. $\mathbf{n}_k\sim\mathcal{CN}(0,\sigma_k^{2}\mathbf{I}_{N_k})$ is the additive white Gaussian noise (AWGN) with zero mean and variance $\sigma_k^2$ experienced by Bob $k$. $\mathbf{z}_{r,k}$ represents the distortion noise caused by the joint effects of the nonlinearity of analog-to-digital converters and automatic gain control (AGC),  the AGC noise, and the oscillator phase noise~\cite{Bai2024}. $\mathbf{n}_r\sim\mathcal{CN}(0,\sigma_r^{2}\mathbf{I}_M)$ stands for noise at RIS, which is related to input and intrinsic devices. $\mathbf{T}=\text{diag}(\mathbf{t})\in\mathbb{C}^{N\times N}$ represents the matrix related to the positions of MAs, which can be expressed as
\begin{equation}
	\label{eq4}
	\mathbf{T}=\begin{bmatrix}{t}_0&\cdots&{0}\\\vdots&\ddots&\vdots\\{0}&\cdots&{t}_{N-1}\end{bmatrix},
\end{equation}
where ${t}_n\in \{0,1\}$ $(n\in\mathcal{N})$ is a binary number used to indicate whether the $n$-th candidate position is selected as a MA position, $\mathbf{t}=[t_0,\ldots,t_{N-1}]^T$. When $t_n=0$, the path gain of the corresponding $n$-th candidate position is zero, excluding it from being a MA position; otherwise, it is considered a valid candidate. The $n$-th candidate position can be mapped to position $\mathbf{t}_n=[n_h,n_v]^T$, given by
\begin{align}
	\label{eqw}
	n_h = \lfloor\frac{n}{N_z}\rfloor, n_v = n-N_z\lfloor\frac{n}{N_z}\rfloor, n_h\in\mathrm{N}_h, n_v\in\mathrm{N}_v.
\end{align}
Assuming the absence of receiver HWIs at Eve, the $k$-th signal received at Eve can be modeled as
\begin{align}
	\label{eq5}
	\mathbf{y}_{e,k}&=(\mathbf{H}_e^H+\mathbf{F}_e^H\boldsymbol{\Theta}\mathbf{G})\mathbf{T}\mathbf{x}+\mathbf n_e+\mathbf{F}_e^H\boldsymbol{\Theta}\mathbf{n}_r,
\end{align}
where $\mathbf{n}_e\sim\mathcal{CN}(0,\sigma_e^{2}\mathbf{I}_{N_e})$ denotes the AWGN with zero mean and variance $\sigma_e^2$ experienced by Eve.

Let $\mathbf{Q}_k=\mathbf{H}_k^H+\mathbf{F}_k^H\boldsymbol{\Theta}\mathbf{G}$ and $\mathbf{Q}_e= \mathbf{H}_e^H+\mathbf{F}_e^H\boldsymbol{\Theta}\mathbf{G}$ represent the aggregated channels corresponding to Bob and Eve, respectively. Then, the received signal $\hat{y}_k$ corresponding to Bob $k$ can be represented as 
\begin{align}\label{eq6}
	\hat{y}_k&=\mathbf{u}_{k}^H\mathbf {y}_{k}\notag\\
	&=\underbrace{\sqrt{\tfrac{\alpha P_0}{K}}\mathbf{u}_{k}^H\mathbf{Q}_k\mathbf{T}\mathbf{v}_ks_k}_{\text{Useful signal components}}+\underbrace{\sqrt{\tfrac{\alpha P_0}{K}}\sum_{i=0,i\neq k}^{K-1}\mathbf{u}_{k}^H\mathbf{Q}_k\mathbf{T}\mathbf{v}_is_i}_{\text{Multi-user interference components}}\notag\\
	&\quad+\underbrace{\sqrt{\tfrac{(1-\alpha) P_0}{K}}\sum_{i=0}^{K-1}\mathbf{u}_{k}^H\mathbf{Q}_k\mathbf{T}\mathbf{v}_{a,i}z+\mathbf{u}_{k}^H\mathbf{n}_k+\mathbf{u}_{k}^H\mathbf{F}_k^H\boldsymbol{\Theta}\mathbf{n}_r}_{\text{Noise components}}\notag\\
	&\quad+\underbrace{\mathbf{u}_{k}^H\mathbf{Q}_k\mathbf{T}\mathbf{z}_t+\mathbf{u}_{k}^H\mathbf{z}_{r,k}}_{\text{HWIs components}},\notag\\
	&=\widetilde{y}_k+\mathbf{u}_{k}^H\mathbf{z}_{r,k}+\mathbf{u}_{k}^H\mathbf{F}_k^H\boldsymbol{\Theta}\mathbf{n}_r,
	\end{align}
	where $\mathbf{u}_{k}\in\mathbb{C}^{N_k\times1}$ represents the receive beamforming vector corresponding to Bob $k$, $\mathbf{u}_{k}^H\mathbf{u}_{k}=1$. $\widetilde{y}_k$ denotes the undistorted received signal at Bob $k$. $\mathbf{u}_{k}^H\mathbf{z}_{r,k}$ follows the Gaussian distribution with zero mean and variance $\mu_{r}\mathbb{E}|{\widetilde{y}}_k|^2$, i.e., $\mathbf{u}_{k}^H\mathbf{z}_{r,k}\sim\mathcal{CN}(0,\mu_{r}\mathbb{E}|{\widetilde{y}}_k|^2)$, where $\mu_r$ denotes the ratio to the undistorted received signal power. Similarly, the received signal $\hat{y}_{e,k}$ corresponding to Eve can be represented as 
	\begin{align}\label{eq7}
\hat{y}_{e,k}&=\mathbf{u}_{e,k}^H\mathbf {y}_{e,k}\notag\\
&=\underbrace{\mathbf{u}_{e,k}^H\mathbf{Q}_e\mathbf{T}(\sqrt{\tfrac{\alpha P_0}{K}}\mathbf{v}_ks_k+\sqrt{\tfrac{(1-\alpha) P_0}{K}}\mathbf{v}_{a,k}z)}_{\text{Useful signals after AN degradation}}\notag\\
&\quad+\sum_{i=0,i\neq k}^{K-1}\mathbf{u}_{e,k}^H\mathbf{Q}_e\mathbf{T}(\sqrt{\tfrac{\alpha P_0}{K}}\mathbf{v}_is_i+\sqrt{\tfrac{(1-\alpha) P_0}{K}}\mathbf{v}_{a,i}z)\notag\\
&\quad+\mathbf{u}_{e,k}^H\mathbf{n}_e+\mathbf{u}_{e,k}^H\mathbf{F}_e^H\boldsymbol{\Theta}\mathbf{n}_r+\mathbf{u}_{e,k}^H\mathbf{Q}_e\mathbf{T}\mathbf{z}_t\notag\\
&=\widetilde{y}_{e,k}+\mathbf{u}_{e,k}^H\mathbf{F}_e^H\boldsymbol{\Theta}\mathbf{n}_r.
\end{align}
where $\widetilde{y}_{e,k}$ denotes the undistorted received signal at Eve, $\mathbf{u}_{e,k}^H\mathbf{u}_{e,k}=1$. $\mathbf{u}_{e,k}\in\mathbb{C}^{N_e\times1}$ denotes the receive beamforming vector corresponding to Eve. Hence, the achievable sum rate of Bob and Eve are respectively given by
\begin{align}
R_U&=\sum_{k=0}^{K-1}\log_2(1+\frac{t_{k}}{\sum_{i=1}^{4}a_{k,i}}),\\
R_E&=\sum_{k=0}^{K-1}\log_2(1+\frac{t_{e,k}}{\sum_{i=1}^{3}b_{k,i}}),
\end{align}
where $a_{k,i}\ (i = 1,2,3,4)$ represents the power of receiver HWIs, multi-user interference (MUI), noise, and transmitter HWIs corresponding to Bob $k$. $t_{k}$ stands for the received useful signal power at Bob $k$. In accordance with \eqref{eq6}, we have
\begin{align}
a_{k,1} &= \tfrac{{\alpha P_0}}{K}\mu_{r}\mathbf{v}_k^H\mathbf{T}^H\mathbf{Q}_k^H\mathbf{U}_k\mathbf{Q}_k\mathbf{T}\mathbf{v}_k,	\\	
a_{k,2} &= \tfrac{{\alpha P_0}}{K}(1+\mu_{r})\sum_{i=0,i\neq k}^{K-1}\mathbf{v}_i^H\mathbf{T}^H\mathbf{Q}_{k}^H\mathbf{U}_k\mathbf{Q}_{k}\mathbf{T}\mathbf{v}_i,
\end{align}
\begin{align}
a_{k,3} &= \tfrac{{(1-\alpha) P_0}}{K}(1+\mu_{r})\sum_{i=0}^{K-1}\mathbf{v}_{a,i}^H\mathbf{T}^H\mathbf{Q}_k^H\mathbf{U}_k\mathbf{Q}_k\mathbf{T}\mathbf{v}_{a,i}\notag\\&\quad+\sigma_r^{2}||\boldsymbol{\Theta}^H\mathbf{F}_{k}\mathbf{U}_k\mathbf{F}_{k}^H\boldsymbol{\Theta}||_F+(1+\mu_{r})\sigma_k^{2},\\
a_{k,4} &= (1+\mu_{r})\mu_{t}\mathbf{u}_{k}^H\mathbf{Q}_{k}\mathbf{T}\mathbf{R}_t\mathbf{T}^H\mathbf{Q}_{k}^H\mathbf{u}_{k},\\
t_{k} &=\tfrac{{\alpha P_0}}{K}\mathbf{v}_k^H\mathbf{T}^H\mathbf{Q}_k^H\mathbf{U}_k\mathbf{Q}_k\mathbf{T}\mathbf{v}_k,
\end{align}
where $\mathbf{U}_k =\mathbf{u}_{k} \mathbf{u}_{k}^H$. Similarly, $b_{k,i}\ (i = 1,2,3)$ denotes the power of MUI, noise, and transmitter HWIs corresponding to Eve. $t_{e,k}$ stands for the received useful signal power at Eve. In accordance with \eqref{eq7}, we have
\begin{align}
b_{k,1} &= \tfrac{{\alpha P_0}}{K}\sum_{i=0,i\neq k}^{K-1}\mathbf{w}_i^H\mathbf{T}^H\mathbf{Q}_{e}^H\mathbf{U}_{e,k}\mathbf{Q}_{e}\mathbf{T}\mathbf{w}_i,\\
b_{k,2} &=\sigma_r^{2}||\boldsymbol{\Theta}^H\mathbf{F}_{e}\mathbf{U}_{e,k}\mathbf{F}_{e}^H\boldsymbol{\Theta}||_F+\sigma_e^{2},\\
b_{k,3}&=\mu_{t}\mathbf{u}_{e,k}^H\mathbf{Q}_{e}\mathbf{T}\mathbf{R}_t\mathbf{T}^H\mathbf{Q}_{e}^H\mathbf{u}_{e,k},\\
t_{e,k} &={\mathbf{w}}_k^H\mathbf{T}^H\mathbf{Q}_e^H\mathbf{U}_{e,k}\mathbf{Q}_e\mathbf{T}{\mathbf{w}}_k,
\end{align}
where ${\mathbf{w}}_k=\sqrt{\tfrac{\alpha P_0}{K}}\mathbf{v}_k+\sqrt{\tfrac{(1-\alpha) P_0}{K}}\mathbf{v}_{a,k}zs_k^*$ and $\mathbf{U}_{e,k}=\mathbf{u}_{e,k}\mathbf{u}_{e,k}^H$. Then, the secrecy sum rate (SSR) is given by
\begin{equation}\label{eq20}
R_s=[R_U-R_E]^{+}.
\end{equation}
\subsection{Channel Model}

The region spanning from $0.62\sqrt{A^3/\lambda}$ to $2A^2/\lambda$ is identified as the radiating NF, commonly referred to as the Fresnel zone, where $\lambda$ denotes the wavelength. In MA systems, the movement of antennas may affect the array aperture $A$, thereby causing the NF region boundaries to vary with antenna displacement, which is not considered in~\cite{Ding2025a} and~\cite{Zhu2025a}. According to (9) in~\cite{Shen2023}, as long as either BS or Bob is in NF relative to the RIS, it can be considered as NF communication. Therefore, deploying the RIS in proximity to Bob ensures the reliability of the NF channel model.

We consider a channel model with imperfect CSI composed of the estimated channel and the estimation error. Here, the estimated channels are deterministic components determined by the azimuth angle, elevation angle, and distance, while the estimation errors are limited by the upper bounds~\cite{Yao2023}. Let ${\mathbf{H}}_k^{c}=\text{diag}(\mathbf{u}_k^H\mathbf{F}_k^H)\mathbf{G}$ and ${\mathbf{H}}_e^c=\text{diag}(\mathbf{u}_{e,k}^H\mathbf{F}_e^H)\mathbf{G}$ denote the estimated cascaded channel corresponding to Bob $k$ and Eve, respectively. Then, given upper bounds $\varepsilon_k$ and $\varepsilon_e$, cascaded channels for Bob $k$ and Eve can be denoted as
\begin{equation}\label{eq21}
{\mathbf{H}}_k^c=\hat{\mathbf{H}}_k^c+\Delta{\mathbf{H}}_k^c,\  \|\Delta{\mathbf{H}}_k^c\|_F\leq\varepsilon_k,
\end{equation}
\begin{equation}\label{eq22}
{\mathbf{H}}_e^c=\hat{\mathbf{H}}_e^c+\Delta{\mathbf{H}}_e^c,\  \|\Delta{\mathbf{H}}_e^c\|_F\leq\varepsilon_e.
\end{equation}
Similarly, given the upper bounds $\hat{\varepsilon}_k$ and $\hat{\varepsilon}_e$ of the direct channel corresponding to Bob $k$ and Eve, we have
\begin{equation}
{\mathbf{h}}_k^d=\mathbf{u}_k^H{\mathbf{H}}_k=\mathbf{u}_k^H\hat{\mathbf{H}}_k+\mathbf{u}_k^H\Delta{\mathbf{H}}_k, \|\mathbf{u}_k^H\Delta{\mathbf{H}}_k\|\leq\hat{\varepsilon}_k,
\end{equation}
\begin{equation}
{\mathbf{h}}_e^d=\mathbf{u}_{e,k}^H{\mathbf{H}}_e=\mathbf{u}_{e,k}^H\hat{\mathbf{H}}_e+\mathbf{u}_{e,k}^H\Delta{\mathbf{H}}_e, \|\mathbf{u}_{e,k}^H\Delta{\mathbf{H}}_e\|\leq\hat{\varepsilon}_e.
\end{equation}

Let $\bar{\phi}$/$\bar{\theta}$, $\bar{\psi}_k$/$\bar{\varphi}_k$, and $\bar{\phi}_k$/$\bar{\theta}_k$ represents the azimuth/elevation angles of departure (AoD) from BS to RIS, from RIS to Bob $k$, and from BS to Bob $k$, respectively. The azimuth/elevation angles of arrival (AoA) from BS to RIS, from RIS to Bob $k$, and from BS to Bob $k$ can be denoted as $\hat{\phi}$/$\hat{\theta}$, $\hat{\psi}_k$/$\hat{\varphi}_k$, and $\hat{\phi}_k$/$\hat{\theta}_k$, respectively. The path-loss coefficients from BS to RIS, from RIS to Bob $k$, and   from BS to Bob $k$ are given by $\gamma_{_{G}}=\frac{\lambda}{\sqrt{4\pi d_G}}$, $\gamma_{f,k}=\frac{\lambda}{\sqrt{4\pi d_{f,k}}}$, and $\gamma_{_{h,k}}=\frac{\lambda}{\sqrt{4\pi d_{h,k}}}$, where $d_{G}$, $d_{{f,k}}$,  and $d_{{h,k}}$ are the distances from BS to RIS, from RIS to Bob $k$, and from BS to Bob $k$. According to geometric characteristics, the horizontal and vertical steering vectors for the channel from BS to RIS can be represented as
\begin{align}
\mathbf{a}(\bar{\phi},\bar{\theta},d_{_{G}})&=\bigg[ e^{-j\frac{2\pi f}{c}\hat{r}_0}, e^{-j\frac{2\pi f}{c}\hat{r}_1},\ldots, e^{-j\frac{2\pi f}{c}\hat{r}_{N_h}}\bigg] ^{T},\\
\mathbf{b}(\bar{\phi},\bar{\theta},d_{_{G}})&= \bigg[ e^{-j\frac{2\pi f}{c}\tilde{r}_0}, e^{-j\frac{2\pi f}{c}\tilde{r}_1},\ldots, e^{-j\frac{2\pi f}{c}\tilde{r}_{N_v}}\bigg] ^{T},
\end{align}
where $\hat{r}_h=r_h-r_l\ (h\in\mathcal N_h)$ and $\hat{r}_v=r_v-r_l\ (v\in\mathcal N_v)$. $r_l$ stands for the distance from MA array to RIS. $r_h=\sqrt{r_l^2+h^2d_{\min}^2-2r_lhd_{\min}\cos\bar{\theta}}$ and $r_v=\sqrt{r_l^2+v^2d_{\min}^2-2r_lvd_{\min}\sin\bar{\theta}\sin\bar{\phi}}$ represent the distances from $h$-th horizontal and $v$-th vertical candidate positions to RIS, respectively. $f$ and $c$ denote frequency and speed of light, respectively. Then, we have the deterministic component $\hat{\mathbf{G}}$, i.e.,
\begin{equation}
\begin{split}
	\hat{\mathbf{G}}=\gamma_{_{G}}e^{-j\frac{2\pi f}{c}r_l}&\big(\mathbf{a}(\bar{\phi},\bar{\theta},d_{_{G}})\otimes\mathbf{b}(\bar{\phi},\bar{\theta},d_{_{G}})\big)\\
	&\big(\hat{\mathbf{a}}(\hat{\phi},\hat{\theta},d_{_{G}})\otimes\hat{\mathbf{b}}(\hat{\phi},\hat{\theta},d_{_{G}})\big).
\end{split}
\end{equation}
$\hat{\mathbf{a}}(\hat{\phi},\hat{\theta},d_{_{G}})$ and $\hat{\mathbf{b}}(\hat{\phi},\hat{\theta},d_{_{G}})$ are given as
\begin{align}
\hat{\mathbf{a}}(\hat{\phi},\hat{\theta},d_{_{G}})&=\bigg[e^{-j\frac{2\pi f}{c}\hat{r}_0^{\prime}}, e^{-j\frac{2\pi f}{c}\hat{r}_1^{\prime}},\ldots, e^{-j\frac{2\pi f}{c}\hat{r}_{M_h}^{\prime}}\bigg] ^{T},\\
\hat{\mathbf{b}}(\hat{\phi},\hat{\theta},d_{_{G}})&= \bigg[ e^{-j\frac{2\pi f}{c}\tilde{r}_0^{\prime}}, e^{-j\frac{2\pi f}{c}\tilde{r}_1^{\prime}},\ldots, e^{-j\frac{2\pi f}{c}\tilde{r}_{M_v}^{\prime}}\bigg] ^{T},
\end{align}
where $\hat{r}_{\hat{h}}^{\prime}=r_{\hat{h}}-r_l\ ({\hat{h}}\in\mathcal M_h)$ and $\hat{r}_{\hat{v}}=r_{\hat{v}}^{\prime}-r_l\ ({\hat{v}}\in\mathcal M_v)$. $r_l$ stands for the distance from MA array to RIS. $r_{\hat{h}}=\sqrt{r_l^2+{\hat{h}}^2d_{\min}^2-2r_l{\hat{h}}d_{\min}\cos\hat{\theta}}$ and $r_{\hat{v}}=\sqrt{r_l^2+{\hat{v}}^2d_{\min}^2-2r_l{\hat{v}}d_{\min}\sin\hat{\theta}\sin\hat{\phi}}$ represent the distances from ${\hat{h}}$-th horizontal and ${\hat{v}}$-th RIS units to MA array, respectively. Similarly, deterministic components $\hat{\mathbf{F}}_k$, $\hat{\mathbf{H}}_k$, $\hat{\mathbf{F}}_e$ and $\hat{\mathbf{H}}_e$ can be obtained.
	Then, the deterministic components of the cascaded channels can be expressed as $\hat{\mathbf{H}}_k^{c}=\text{diag}(\mathbf{u}_k^H\hat{\mathbf{F}}_k^H)\hat{\mathbf{G}}$ and $\hat{\mathbf{H}}_e^c=\text{diag}(\mathbf{u}_{e,k}^H\hat{\mathbf{F}}_e^H)\hat{\mathbf{G}}$.
	
	\subsection{Problem Formulation}
	In this paper, we aim to maximize SSR by jointly optimizing the transmit beamforming vectors $\mathbf{v}_{a,k}$ and $\mathbf{v}_k$, the power allocation factor $\alpha$, the MA position selection matrix $\mathbf{T}$, the phase shift matrix $\boldsymbol{\Theta}$, and the receive beamforming vectors $\mathbf{u}_k$ and $\mathbf{u}_{e,k}$. The SSR maximization problem can be formulated as
	\begin{subequations}
		\begin{align}
			\label{eq34a}
			\text{P1:}\quad\quad&\underset{\mathbf{v}_{a,k},\mathbf{v}_k,\alpha,\boldsymbol{\Theta},\mathbf{T},\mathbf{u}_k,\mathbf{u}_{e,k}}{\max}\ R_s
			\\ \label{eq34b}
			\mathrm{s.t.}\  &\text{C1}:\mathbf{v}_k^H\mathbf{T}\mathbf{v}_k = 1, \forall k\\\label{eq34c}
			&\text{C2}:\mathbf{v}_{a,k}^H\mathbf{T}\mathbf{v}_{a,k} =1, \forall k\\ 
			&\text{C3}:\mathbf{u}_k^H\mathbf{u}_k=1,\mathbf{u}_{e,k}^H\mathbf{u}_{e,k}=1,\\
			&\text{C4}:\mathbf{u}_{e,k}^H\mathbf{Q}_e\mathbf{T}\mathbf{w}_k=0, \forall k,\\
			&\text{C5}:\mathbf{u}_{k}^H\mathbf{Q}_k\mathbf{T}\mathbf{v}_{a,i}=0, \forall k,i\in\mathcal{K},\\
			&\text{C6}:\|\mathbf{T}\|_0=N_a, t_n \in \{0,1\}, \forall n,\\\label{34h}
			&\text{C7}:\|\mathbf{t}_{n_a}-\mathbf{t}_{n_a^\prime}\|\geq d, \notag\\
			&\qquad n_a, n_a^\prime\in\{n|t_n=1\},n_a\neq n_a^\prime,\\\label{34i}
			&\text{C8}:\sum_{k=0}^{K-1}\mathbf{w}_k^H\mathbf{T}^H\mathbf{G}^H\boldsymbol{\Theta}^H\boldsymbol{\Theta}\mathbf{G}\mathbf{T}\mathbf{w}_k+\sigma_r^2\bm{\theta}^H\bm{\theta}\leq P_{\text{RIS}},
		\end{align}
	\end{subequations}
	where $d$ represents the minimum distance between MAs limited by antenna size. $P_{\text{RIS}}$ denotes the reflection power threshold. In P1, we aim to maximize SSR, and the BS needs to dissipate all the transmission power $P_0$, i.e., constraints C1 and C2. For fairness, as shown in constraint C3, the receive beamforming vectors are normalized. In the system, AN is introduced to reduce the received power and add noises at Eve while not interfering with Bob, so we have constraints C4 and C5. Constraints C6 and C7 limit the MR and minimum spacing of MAs. C8 represents the power budget of active RIS. 
	
	Due to variable coupling effects and the discrete nature of MA positioning, P1 becomes a non-convex optimization challenge that resists efficient solutions. While the exhaustive search could theoretically solve this problem, its prohibitive computational complexity renders it impractical. For instance, selecting optimal positions for 60 MAs from 1000 candidate locations would require evaluating $\tfrac{1000!}{60!940!}$ combinations.

	\section{Proposed Optimization Algorithms}\label{sec:4}
	In this section, we propose three novel optimization algorithms to address the design challenges. Initially, we develop an iterative optimization algorithm based on phase alignment and leakage theory to jointly design the beamforming vector and RIS phase shift matrix under HWIs and imperfect CSI. Subsequently, two distinct MA position optimization approaches are proposed: a uniform sampling-based method and a compressive sensing (CS)-based non-uniform sampling technique. The proposed methods achieve low computational complexity through group-based optimization of the candidate positions. 
	\subsection{Problem Reformulation}
	According to \eqref{eq21} and \eqref{eq22}, the channels corresponding to Bob $k$ and Eve under all uncertainties can be written as
	\begin{equation}\label{eq35}
		\mathbf{A}_k = 
		\begin{bmatrix}\mathbf{H}_k^c\\\mathbf{h}_k^d
		\end{bmatrix}=
		\begin{bmatrix}\hat{\mathbf{H}}_k^c\\\mathbf{u}_k^H\hat{\mathbf{H}}_k
		\end{bmatrix}+
		\begin{bmatrix}\Delta{\mathbf{H}}_k^c\\\mathbf{u}_k^H\Delta{\mathbf{H}}_k
		\end{bmatrix}=
		\hat{\mathbf{A}}_k+\Delta{\mathbf{A}}_k,
	\end{equation}
	\begin{equation}\label{eq36}
		\mathbf{A}_e = 
		\begin{bmatrix}\mathbf{H}_e^c\\\mathbf{h}_e^d
		\end{bmatrix}=
		\begin{bmatrix}\hat{\mathbf{H}}_{e}^c\\\mathbf{u}_{e,k}^H\hat{\mathbf{H}}_e
		\end{bmatrix}+
		\begin{bmatrix}\Delta{\mathbf{H}}_e^c\\\mathbf{u}_{e,k}^H\Delta{\mathbf{H}}_e
		\end{bmatrix}=
		\hat{\mathbf{A}}_e+\Delta{\mathbf{A}}_e,
	\end{equation}
	where $\|\Delta{\mathbf{A}}_k\|_F\leq\tilde{\varepsilon}_k=\sqrt{{\varepsilon}_k^2+\hat{\varepsilon}_k^2}$ and $\|\Delta{\mathbf{A}}_e\|_F\leq\tilde{\varepsilon}_e=\sqrt{{\varepsilon}_e^2+\hat{\varepsilon}_e^2}$. Let $\hat{\bm{\theta}}=[\bm{\theta}^T,1]\in\mathrm{C}^{1\times(M+1)}$, constraints C4 and C5 can be equivalently expressed as
	\begin{align}
		\text{C4}^\prime:\ &\hat{\bm{\theta}}\mathbf{A}_e\mathbf{T}\mathbf{w}_k=0, \forall k,\\\label{eq38}
		\text{C5}^\prime:\ &\hat{\bm{\theta}}\mathbf{A}_k\mathbf{T}\mathbf{v}_{a,i}=0, \forall k, i\in\mathcal{K},
	\end{align}
	respectively. Under imperfect CSI conditions, $\text{C4}^\prime$ and $\text{C5}^\prime$ may fail. We perform beamforming design using the estimated channel (i.e., the deterministic components) and take the upper bound of the channel estimation error, given by
	\begin{align}\label{eq39}
		\hat{\bm{\theta}}\hat{\mathbf{A}}_e\mathbf{T}\mathbf{w}_k&=0,\\\label{eq40}
		\hat{\bm{\theta}}\hat{\mathbf{A}}_k\mathbf{T}\mathbf{v}_{a,i}&=0,i\in\mathcal{K}.
	\end{align}
	With \eqref{eq39} and \eqref{eq40}, we can obtain
	\begin{align}\label{41}
		\|\hat{\bm{\theta}}\hat{\mathbf{A}}_e\mathbf{T}\mathbf{w}_k\|&
		\leq\hat{\varepsilon}_e\|\hat{\bm{\theta}}\|\|\mathbf{T}\mathbf{w}_k\|\notag\\&=\hat{\varepsilon}_e\tfrac{P_0}{K}\|\hat{\bm{\theta}}\|,\\\label{42}
		\|\sum_{i=0}^{K-1}\hat{\bm{\theta}}\hat{\mathbf{A}}_k\mathbf{T}\mathbf{v}_{a,i}\|&
		\leq\sum_{i=0}^{K-1}\hat{\varepsilon}_k\|\hat{\bm{\theta}}\|\|\mathbf{T}\mathbf{v}_{a,i}\|\notag\\&=\hat{\varepsilon}_k{(1-\alpha)P_0}\|\hat{\bm{\theta}}\|.
	\end{align}
	Therefore, constraints $\text{C4}^\prime$ and $\text{C5}^\prime$ can be expressed as
	\begin{equation}
		\begin{split}
			\label{eq43}&\text{C9}: \eqref{eq39},\text{C10}:\eqref{eq40},\\&\text{C11}:\underset{\bm{\theta}}{\min}\ (\hat{\varepsilon}_e\tfrac{P_0}{K}+\hat{\varepsilon}_k{(1-\alpha)P_0})\|\hat{\bm{\theta}}\|,
		\end{split} 
	\end{equation}
	which indicates that reducing the RIS reflection power, saving BS transmission power, and enhancing energy allocation to CM transmission can help alleviate the degradation caused by imperfect CSI. Here, $\alpha$ should be maximized to mitigate the impact of imperfect CSI.  When $\alpha=1$, there is no AN, which means that the entire transmission power is used to enhance the transmission rate of Bob.  When $0<\alpha<1$, maximizing $\alpha$ is equivalent to maximizing the proportion of transmit power used for CM transmission. When $\alpha=0$, there is no CM transmission.
	
	By substituting \eqref{eq34b} and \eqref{eq34c} into $\mathbf{w}_k$, we can obtain
	\begin{equation}\label{eq44}
		\text{C12}:\mathbf{w}_k^H\mathbf{T}\mathbf{w}_k=\tfrac{P_0}{K}, \forall k.
	\end{equation}
	Therefore, the optimization problem can be reformulated as 
	\begin{subequations}
		\begin{align}
			\label{eq46a}
			\text{P2:}&\quad\underset{\mathbf{w}_k,\bm{\theta},\mathbf{T},\mathbf{u}_k,\mathbf{u}_{e,k}}{\max}\ \ \ R_s(\mathbf{w}_k,\bm{\theta},\mathbf{T},\mathbf{u}_k,\mathbf{u}_{e,k})
			\\ \label{eq46b}
			&\quad\qquad\mathrm{s.t.}\  \text{C3}, \text{C6}, \text{C7}, \text{C8}, \text{C9}, \text{C10}, \text{C11}, \text{C12}.
		\end{align}
	\end{subequations}
	\subsection{Optimize $\mathbf{w}_k$ and $\bm{\Theta}$}
	To address P2, an iterative optimization method for $\mathbf{w}_k$ and $\bm{\Theta}$ is first proposed. Define the two initial states $\boldsymbol{\Theta}^{0}$ of $\boldsymbol{\Theta}$ as
	\begin{equation}\label{eq47}
		\begin{split}
			\boldsymbol{\Theta}^0=\left\lbrace
			\begin{aligned}
				\mathbf{0}_M,&\ \text{no RIS},\\
				\mathbf{I}_M,&\ \text{passive RIS}.
			\end{aligned}\right. 
		\end{split}
	\end{equation}	
	Also, we set the initial state $\mathbf{T}^0$ of $\mathbf{T}$ as
	\begin{equation}\label{eq48}
		\begin{split}
			\mathbf{T}^0=\left\lbrace
			\begin{aligned}
				&\begin{bmatrix}\mathbf{I}_{N_a}&\mathbf{0}\\\mathbf{0}&\mathbf{0}
				\end{bmatrix},\ \text{FPA with $N_a$ antennas,}\\
				&\begin{bmatrix}\mathbf{I}_{K+1}&\mathbf{0}\\\mathbf{0}&\mathbf{0}
				\end{bmatrix},\ \text{FPA with $K+1$ antennas,}\\
				&\ \mathbf{I}_{N},\ \text{FPA with $N$ antennas.}
			\end{aligned}\right. 
		\end{split}
	\end{equation}
	where the minimum number of MAs required by the system is $K+1$. In iterative optimization methods, due to the coupling of multiple variables, it is generally difficult to determine the optimal initial values. With the initialization settings illustrated in \eqref{eq47} and \eqref{eq48}, it is convenient to subsequently analyze the performance gains brought by MA array. The initial values of 
	$\mathbf{U}_{k}$ and $\mathbf{U}_{e,k}$ are set as identity matrices, representing equal-power reception.
	
	Then, given $\mathbf{u}_k$, $\mathbf{u}_{e,k}$, $\boldsymbol{\Theta}$, and $\mathbf{T}$, according to the leakage theory, we have signal-to-leakage-noise ratio (SLNR) with respect to (w.r.t.) $\mathbf{v}_k$, i.e.,
	\begin{align}\label{eq49}
		\begin{split}
			\mathrm{SLNR}_{\mathbf{v}_k} &= \frac{\mathbf{v}_k^H\mathbf{T}^H\mathbf{L}_{1,k}\mathbf{T}\mathbf{v}_k}{\mathbf{v}_k^H\mathbf{T}^H(\hat{\mathbf{A}}_e^H\hat{\bm{\theta}}^H\hat{\bm{\theta}}\hat{\mathbf{A}}_e+\mathbf{L}_{2,k}+\mathbf{L}_{3,k})\mathbf{T}\mathbf{v}_k},
		\end{split}
	\end{align}
	where
	\begin{align}
		\mathbf{L}_{1,k} &= \mathbf{Q}_k^H\mathbf{U}_{k}\mathbf{Q}_k,\\
		\mathbf{L}_{2,k} &= \sum_{i=0,i\neq k}^{K-1}\mathbf{Q}_{i}^H\mathbf{U}_{i}\mathbf{Q}_{i},\\
		\mathbf{L}_{3,k} &= (\sigma_k^2+\sigma_r^{2}\|\boldsymbol{\Theta}^H\mathbf{F}_{k}\mathbf{U}_{k}\mathbf{F}_{k}^H\boldsymbol{\Theta}\|_F)\mathbf{I}_N.
	\end{align}
	The meaning of \eqref{eq49} is to minimize the energy leaked to other Bob and Eve, which is equivalent to the SSR maximization problem under perfect CSI and HWI-free conditions. By substituting \eqref{eq39} and \eqref{eq40} into \eqref{eq49}, the SLNR w.r.t. $\mathbf{w}_k$ can be represented as
	\begin{align}\label{eq52}
		\begin{split}
			\mathrm{SLNR}_{\mathbf{w}_k} &= \frac{\mathbf{w}_k^H\mathbf{T}^H\mathbf{L}_{1,k}\mathbf{T}\mathbf{w}_k}{\mathbf{w}_k^H\mathbf{T}^H(\mathbf{L}_{2,k}+\tfrac{K}{P_0}\mathbf{L}_{3,k})\mathbf{T}\mathbf{w}_k}.
		\end{split}
	\end{align}
	Considering imperfect CSI, according to \eqref{eq43}, \eqref{eq52} can be denoted as
	\begin{equation}\label{eq53}
		\mathrm{SLNR}_{\mathbf{w}_k}^\prime = \frac{\mathbf{w}_k^H\mathbf{T}^H\mathbf{L}_{1,k}\mathbf{T}\mathbf{w}_k}{\mathbf{w}_k^H\mathbf{T}^H\mathbf{L}_{4,k}\mathbf{T}\mathbf{w}_k},
	\end{equation}
	where $\mathbf{L}_{4,k}=\mathbf{L}_{2,k}+\tfrac{K}{P_0}\mathbf{L}_{3,k}+(\hat{\varepsilon}_e\tfrac{P_0}{K}+\hat{\varepsilon}_k{(1-\alpha)P_0})\|\hat{\bm{\theta}}\|\mathbf{I}_N$. In practice, the HWIs at Eve are difficult to obtain and may be much smaller than Bob. With \eqref{eq39} and \eqref{41}, under the HWIs-free conditions, the upper bound of the achievable sum rate corresponding to Eve can be derived as 
	\begin{equation}
		R_E\leq\sum_{k=0}^{K-1}\log_2\big(1+\frac{\hat{\varepsilon}_e{P_0}\|\hat{\bm{\theta}}\|}{{K}b_{k,2}}\big).
	\end{equation}
	
	In the presence of HWIs, a first-order Taylor approximation of $a_{k,4}$ yields
	\begin{equation}\label{eq54}
		a_{k,4}= \sum_{i=1}^{K-1}\mathbf{w}_i^H\widetilde{\text{diag}}(\mathbf{T}^H\mathbf{Q}_{k}^H\mathbf{U}_{k}\mathbf{Q}_{k}\mathbf{T})\mathbf{w}_i.\\
	\end{equation}
	Subsequently, we define the covariance matrix $\mathbf{L}_{5,k}$ of the virtual channel characterizing HWIs, denoted as
	\begin{equation}\label{eq55}
		\begin{split}
			\mathbf{L}_{5,k}=& (1+\mu_r)\mu_t\sum_{i=1}^{K-1}\widetilde{\text{diag}}(\mathbf{T}^H\mathbf{Q}_{i}^H\mathbf{U}_{i}\mathbf{Q}_{i}\mathbf{T})\\&+\mu_r(\mathbf{L}_{1,k}+\mathbf{L}_{2,k}+\tfrac{K\sigma_k}{P_0}^2\mathbf{I}_N).
		\end{split}
	\end{equation}
	Thus, the SSR $R_s(\mathbf{w}_k)$ w.r.t. $\mathbf{w}_k$ can be expressed as
	\begin{align}\label{eq56}
		\begin{split}
			R_s(\mathbf{w}_k)&\triangleq \frac{\mathbf{w}_k^H\mathbf{T}^H\mathbf{L}_{1,k}\mathbf{T}\mathbf{w}_k}{\mathbf{w}_k^H\mathbf{T}^H(\mathbf{L}_{4,k}+\mathbf{L}_{5,k})\mathbf{T}\mathbf{w}_k}.
		\end{split}
	\end{align}
	Based on the above analysis, the problem w.r.t. $\mathbf{w}_k$ can be expressed as \begin{subequations}\label{eq57}
		\begin{align}
			\label{eq57a}
			\text{P3:}&\quad\underset{\mathbf{w}_k}{\max}\ \frac{\mathbf{w}_k^H\mathbf{T}^H\mathbf{L}_{1,k}\mathbf{T}\mathbf{w}_k}{\mathbf{w}_k^H\mathbf{T}^H(\mathbf{L}_{4,k}+\mathbf{L}_{5,k})\mathbf{T}\mathbf{w}_k}
			\\ \label{eq57b}
			&\quad\mathrm{s.t.}\  \text{C8}, \text{C9}, \text{C12}.
		\end{align}
	\end{subequations}
	In \eqref{eq57}, by constructing the upper bound of channel estimation error under imperfect CSI and virtual channels related to HWIs, a SLNR mathematical model under imperfect CSI and HWIs is derived to replace the SSR maximization problem w.r.t. $\mathbf{w}_k$. 
	
	By substituting \eqref{eq47} into \eqref{34i}, it can be observed that constraint C8 is satisfied in no RIS scenarios. For passive RIS-aided scenarios, we have
	\begin{equation}\label{eq58}
		\begin{split}
			&\quad\sum_{k=0}^{K-1}\mathbf{w}_k^H\mathbf{T}^H\mathbf{G}^H\boldsymbol{\Theta}^H\boldsymbol{\Theta}\mathbf{G}\mathbf{T}\mathbf{w}_k+\sigma_r^2\bm{\theta}^H\bm{\theta}\\&\leq\sum_{k=0}^{K-1} \|\mathbf{G}\mathbf{T}\|_F^2\|\bm{\theta}\|^2\|\mathbf{w}_k\|^2+\sigma_r^2\|\bm{\theta}\|^2\\&= ({P_0}\|\mathbf{G}\mathbf{T}\|_F^2+\sigma_r^2)M\\&\leq P_{\text{RIS}}.
		\end{split}
	\end{equation}
	Therefore, for the initial states, C8 can be removed, and we have
	\begin{subequations}\label{eq59}
		\begin{align}
			\label{eq59a}
			\text{P4:}&\quad\underset{\mathbf{w}_k}{\max}\ \frac{\mathbf{w}_k^H\mathbf{T}^H\mathbf{L}_{1,k}\mathbf{T}\mathbf{w}_k}{\mathbf{w}_k^H\mathbf{T}^H(\mathbf{L}_{4,k}+\mathbf{L}_{5,k})\mathbf{T}\mathbf{w}_k}
			\\ \label{eq59b}
			&\quad\mathrm{s.t.}\  \text{C9}, \text{C12}.
		\end{align}
	\end{subequations}
	P4 can be solved by exploiting Rayleigh-Ritz theorem and null-space projection. Specifically, auxiliary variable $\bm{\Gamma}_1=\mathbf{I}_{N}-\mathbf{T}^H\hat{\mathbf{A}}_e^H\hat{\bm{\theta}}^H(\hat{\bm{\theta}}\hat{\mathbf{A}}_e\mathbf{T}\hat{\mathbf{A}}_e^H\hat{\bm{\theta}}^H)^{-1}\hat{\bm{\theta}}\hat{\mathbf{A}}_e\mathbf{T}$ is introduced and $\mathbf{w}_k=\sqrt{\tfrac{P_0}{K}}\bm{\Gamma}_1\bm{\rho}_k$ is substituted into P4 to obtain
	\begin{equation}\label{eq60}
		\begin{split}
			\text{P5:}&\quad\underset{\bm{\rho}_k}{\max}\ \frac{\bm{\rho}_k^H\bm{\Gamma}_1^H\mathbf{T}^H\mathbf{L}_{1,k}\mathbf{T}\bm{\Gamma}_1\bm{\rho}_k}{\bm{\rho}_k^H\bm{\Gamma}_1^H\mathbf{T}^H(\mathbf{L}_{4,k}+\mathbf{L}_{5,k})\mathbf{T}\bm{\Gamma}_1\bm{\rho}_k}
			\\ 
			&\quad\mathrm{s.t.}\ \text{C13}: \bm{\rho}_k^H\bm{\rho}_k=1.
		\end{split}
	\end{equation}
	P5 is a classic Generalized Rayleigh Quotient (GRQ) problem. The optimal solution for $\bm{\rho}_k$ is the eigenvector associated with the largest eigenvalue of matrix $\mathbf{X}_k$, where
	\begin{align}\label{eq61}
		\mathbf{X}_k&=\mathbf{B}_k^{-\tfrac{1}{2}}\bm{\Gamma}_1^H\mathbf{T}^H\mathbf{L}_{1,k}\mathbf{T}\bm{\Gamma}_1\mathbf{B}_k^{-\tfrac{1}{2}},\\
		\mathbf{B}_k&=\bm{\Gamma}_1^H\mathbf{T}^H(\mathbf{L}_{4,k}+\mathbf{L}_{5,k})\mathbf{T}\bm{\Gamma}_1.
	\end{align}
	Then, $\mathbf{w}_k$ can be calculated exploiting $\bm{\rho}_k$. 
	
	Given $\mathbf{u}_k$, $\mathbf{u}_{e,k}$, $\mathbf{w}_k$, and $\mathbf{T}$, $\boldsymbol{\Theta}$ can be optimized. Here, the initial state of $\boldsymbol{\Theta}$ is updated. Since both the amplitude and phase shift of the active RIS can be adjusted, constraint C8 must be taken into account. Let $\mathbf{H}=[\mathbf{A}_0\mathbf{T}\mathbf{w}_0,\mathbf{A}_1\mathbf{T}\mathbf{w}_1,\ldots,\mathbf{A}_{K-1}\mathbf{T}\mathbf{w}_{K-1}]\in\mathrm{C}^{(M+1)\times K}$ denote the set of all Bob channels. According to \eqref{eq35} and \eqref{eq36}, the SLNR w.r.t. $\hat{\bm{\theta}}$ (i.e., $\boldsymbol{\Theta}$) can be expressed as 
	\begin{subequations}
		\begin{align}
			\text{P6:}&\quad\underset{\hat{\bm{\theta}}}{\max}\ \frac{\hat{\bm{\theta}}\mathbf{L}_{1,k}^\prime\hat{\bm{\theta}}^H}{\hat{\bm{\theta}}(\mathbf{L}_{4,k}^\prime+\mathbf{L}_{5,k}^\prime)\hat{\bm{\theta}}^H}
			\\ 
			&\quad\mathrm{s.t.}\  \text{C8}, \text{C9}.
		\end{align}
	\end{subequations}
	where 
	\begin{align}
		\mathbf{L}_{1,k}^\prime&=\mathbf{H}\mathbf{H}^H,\\
		\mathbf{L}_{4,k}^\prime&=((1+\mu_r)\sigma_k^2+\hat{\varepsilon}_e\tfrac{P_0}{K}+\hat{\varepsilon}_k{(1-\alpha)P_0})\mathbf{I}_{M+1},\\
		\mathbf{L}_{5,k}^\prime&=(1+\mu_r)\sigma_r^{2}\text{diag}(\mathbf{u}_k^H\mathbf{F}_k^H)\text{diag}(\mathbf{F}_k\mathbf{u}_k)+\mu_r\mathbf{L}_{1,k}^\prime\notag\\&
		\quad\ (1+\mu_r)\mu_t\sum_{i=0}^{K-1}\mathbf{A}_i\widetilde{\text{diag}}(\mathbf{T}\sum_{k=0}^{K-1}\mathbf{w}_k\mathbf{w}_k^H)\mathbf{A}_i^H.
	\end{align}
	Problem P6 can be transformed into a Rayleigh quotient form by relaxing constraint C8 and introducing auxiliary variable $\hat{\bm{\theta}}\hat{\bm{\theta}}^H=1$. Then, the normalization method in \cite{Shu2023} can be applied to satisfy constraint C8. However, when the direct link and cascaded link are not perfectly aligned, signal gain will be reduced because the normalization of $\boldsymbol{\Theta}$ only performs scaling for the cascaded link. Furthermore, the normalization operation cannot align the signals to all Bob. To address this issue, we reformulate P6 into an optimization scheme w.r.t $\boldsymbol{\Theta}$, as detailed in the following.
	
	As specified in \eqref{eq49}, \eqref{eq53}, and \eqref{eq56}, to enhance SSR, we aim to optimize $\bm{\theta}$ (i.e., $\boldsymbol{\Theta}$) for enhanced power gain at Bob while minimizing power leakage to the Eve channel, other Bob channels, virtual channel corresponding to HWIs, and imperfect CSI effects. Specifically, an auxiliary variable $\bm{\theta}_d^T=\bm{\theta}^j-\bm{\theta}^{j-1}$ is introduced to represent the difference between the phase shift matrices of the $j$-th and $(j-1)$-th iterations, where $\bm{\theta}^j=\text{diag}(\boldsymbol{\Theta}^j)$. 
	The received power at Bob $k$ in the $j$-th iteration is not less than that in the $(j-1)$-th iteration, we have
	\begin{align}
		\label{eq65}&\text{C14}:\bm{\theta}_d\hat{\mathbf{H}}_k^{c}\mathbf{T}\mathbf{w}_k=\xi_1\bm{\theta}^{j-1}\hat{\mathbf{H}}_k^{c}\mathbf{T}\mathbf{w}_k, \forall k,
	\end{align} 
	where $\xi_1>0$ denotes the proportionality factor. With $\mathbf{w}_k$, the deterministic components of Eve channel can be transformed into a more tractable form and denoted as 
	\begin{equation}\label{eq63}
		\begin{split}
			\bar{\mathbf{a}}_k=(\hat{\mathbf{H}}_e^c+\mathbf{h}_e^d)\mathbf{T}\mathbf{w}_k.
		\end{split}
	\end{equation}
	According to $(\bm{\theta}^j)^T\bar{\mathbf{a}}_k=0$, C9 can be rewritten as
	\begin{align}
		\label{eq68}&\text{C15}:\bm{\theta}_d\bar{\mathbf{a}}_k=0, \forall k.
	\end{align} 
	To minimize multi-user interference, we have
	\begin{equation}
		\begin{split}
			\label{eq69}\text{C16}:&\quad \|\big((\bm{\theta}^{j-1})^T+\bm{\theta}_d\big)\sum_{i=0,i\neq k}^{K-1}\hat{\mathbf{H}}_k^{c}\mathbf{T}\mathbf{w}_i\|^2\\&\leq\xi_2\|(\bm{\theta}^{j-1})^T\sum_{i=0,i\neq k}^{K-1}\hat{\mathbf{H}}_k^{c}\mathbf{T}\mathbf{w}_i\|^2, \forall k,
		\end{split}
	\end{equation} 
	where $0\leq\xi_2\leq(1+\xi_1)^2$. In terms of HWI, $a_{k,4}$ can be rewritten as
	\begin{align}
		\label{eq70}
		\begin{split} a_{k,4}=\|\bm{\theta}^T\hat{\mathbf{H}}_k^{c}\cdot\sum_{i=0}^{K-1}|\mathbf{w}_i^T\mathbf{T}|\|^2.
		\end{split}
	\end{align} 
	Then, the constraint for reducing the impact of HWIs can be given as
	\begin{align}
		\label{eq71}
		\begin{split}
			\text{C17}:&\quad\ \|\big((\bm{\theta}^{j-1})^T+\bm{\theta}_d\big)\hat{\mathbf{H}}_k^{c}\cdot\sum_{i=0}^{K-1}|\mathbf{w}_k^T\mathbf{T}|\|^2\\&\leq\|(\bm{\theta}^{j-1})^T\hat{\mathbf{H}}_k^{c}\cdot\sum_{i=0}^{K-1}|\mathbf{w}_k^T\mathbf{T}|\|^2,\forall k.
		\end{split}
	\end{align} 
	According to the derivation in \eqref{eq43}, $\|\bm{\theta}\|^2$ can be minimized to reduce the influence of imperfect CSI. In order to obtain SSR gain while satisfying constraint C8, we formulate the optimization problem w.r.t. to $\bm{\theta}_d$ as
	\begin{equation}
		\begin{split}\label{eq72}
			\text{P7:}\ &\underset{\bm{\theta}_d, \xi_1}{\min}\ -\xi_1\\ 
			&\mathrm{s.t.}\  \text{C8}, \text{C14}, \text{C15}, \text{C16}, \text{C17},\\
			&\quad\ \ \text{C18}: \xi_1\geq0.
		\end{split}
	\end{equation}
	P7 is a convex optimization problem and can be solved by CVX or YALMIP. Then, $\bm{\Theta}=\text{diag}(\bm{\theta}^{j-1}+\bm{\theta}_d^T)$ can be obtained. The joint optimization of $\mathbf{w}_k$ and $\bm{\Theta}$ is described in Algorithm \ref{alg:alg1}.
	\begin{algorithm}
		\renewcommand{\algorithmicrequire}{\textbf{Input:}}
		\renewcommand{\algorithmicensure}{\textbf{Output:}}
		\caption{The proposed joint optimization method for $\mathbf{w}_k$ and $\bm{\Theta}$}\label{alg:alg1}
		\begin{algorithmic}[1]
			\REQUIRE $P_0$, $P_{\text{RIS}}$, $\mu_r$, $\mu_t$, $\sigma_k^2$, $\sigma_r^2$, $\sigma_e^2$, $N$, $N_a$, $M$, $K$, $\xi_1$, $\epsilon$.
			\ENSURE $\mathbf{w}_k$, $\boldsymbol{\Theta}$.
			\STATE Initialize $\mathbf{u}_k$, $\mathbf{u}_{e,k}$, $\boldsymbol{\Theta}$, and $\mathbf{T}$.
			\STATE \textbf{for} $j = 1:1 :J$ \textbf{do}
			\STATE\hspace{0.5cm} Calculate $\bm{\rho}_k$ according to \eqref{eq60}. \STATE\hspace{0.5cm} Update $\mathbf{w}_k=\sqrt{\tfrac{K}{P_0}}\bm{\Gamma}_1\bm{\rho}_k$.
			\STATE\hspace{0.5cm} Calculate $\xi_2\|(\bm{\theta}^{j-1})^T\sum_{i=0,i\neq k}^{K-1}\hat{\mathbf{H}}_k^{c}\mathbf{T}\mathbf{w}_k\|^2$, $\|(\bm{\theta}^{j-1})^T\hat{\mathbf{H}}_k^{c}\cdot|\mathbf{w}_k^T\mathbf{T}|\|^2$, $\xi_1\bm{\theta}^{j-1}\hat{\mathbf{H}}_k^{c}\mathbf{T}\mathbf{w}_k$, and $\bar{\mathbf{a}}_k$.
			\STATE\hspace{0.5cm} Optimize $\bm{\theta}_d$ according to \eqref{eq72}.		\STATE\hspace{0.5cm} Calculate $\bm{\theta}^j=\bm{\theta}^{j-1}+\bm{\theta}_d^T$.
			\STATE\hspace{0.5cm} Calculate $R_s^{j}$ according to \eqref{eq20}.
			
			\STATE\hspace{0.5cm} \textbf{if} $|R_s^{j}-R_s^{j-1}|\leq\epsilon$ \textbf{then}
			\STATE\hspace{0.5cm}\hspace{0.5cm} Break.
			\STATE\hspace{0.5cm} \textbf{end if}
			\STATE\textbf{end for}
			\STATE Set $\bm{\theta} = \bm{\theta}^{j-1}+\bm{\theta}_d^T$.
			\STATE\textbf{return} $\mathbf{w}_k$, $\boldsymbol{\Theta}$.
		\end{algorithmic}
	\end{algorithm}
	\subsection{Proposed MA Position Optimization Methods}
	Given $\mathbf{u}_k$, $\mathbf{u}_{e,k}$, $\mathbf{w}_k$, and $\boldsymbol{\Theta}$, $\mathbf{T}$ can be optimized. The deployment of RIS provides enhanced DoF for NF communication channels, resulting in higher channel rank. This enables the system to transmit signals with varying qualities to Bob. Capitalizing on this characteristic, we propose two distinct algorithms featuring uniform and non-uniform candidate position selection strategies, respectively.
	\subsubsection{Uniform grouping method}
	The proposed MA position optimization strategy involves dividing the candidate positions into $\lfloor\frac{N}{n_0}\rfloor$ groups that satisfy the minimum spacing requirements, where $n_0$ denotes the number of candidate positions per group. After obtaining $\lfloor\frac{N}{n_0}\rfloor$ SSR values and sorting them, the positions of all the MAs are determined. The $l$-th position matrix that satisfies the minimum spacing $d$ is represented as $\hat{\mathbf{T}}_{l}$, where $l\in\{0,1,\ldots,\lfloor\frac{N}{n_0}\rfloor-1\}$ and $\|\mathbf{\hat{\mathbf{T}}_{l}}\|_0\leq N_a$. Note that for all Bob and Eve, at least $K+1$ MAs are required. Specifically, with $\hat{\bm{\theta}}$ obtained by Algorithm \ref{alg:alg1}, by introducing $\bm{\Gamma}_{2,l}=\mathbf{I}_{N}-\hat{\mathbf{T}}_{l}\hat{\mathbf{A}}_e^H\hat{\bm{\theta}}^H(\hat{\bm{\theta}}\hat{\mathbf{A}}_e\hat{\mathbf{T}}_{l}\hat{\mathbf{A}}_e^H\hat{\bm{\theta}}^H)^{-1}\hat{\bm{\theta}}\hat{\mathbf{A}}_e\hat{\mathbf{T}}_{l}$ and $\mathbf{w}_{k,l}=\sqrt{\tfrac{P_0}{K}}\bm{\Gamma}_{2,l}\hat{\bm{\rho}}_{k,l}$, the optimization w.r.t. $\mathbf{w}_{k,l}$ can be formulated as
	\begin{subequations}\label{eq73}
		\begin{align}
			\text{P8:}\ \underset{\hat{\bm{\rho}}_{k,l}}{\max}&\ \frac{\hat{\bm{\rho}}_{k,l}^H\bm{\Gamma}_{2,l}^H\hat{\mathbf{T}}_{l}\mathbf{L}_{1,k}\bm{\Gamma}_{2,l}\hat{\bm{\rho}}_{k,l}}{\hat{\bm{\rho}}_{k,l}^H\bm{\Gamma}_{2,l}^H\hat{\mathbf{T}}_{l}(\mathbf{L}_{4,k}+\mathbf{L}_{5,k})\bm{\Gamma}_{2,l}\hat{\bm{\rho}}_{k,l}}
			\\
			\mathrm{s.t.}\  &\text{C19}: \hat{\bm{\rho}}_{k,l}^H\hat{\bm{\rho}}_{k,l}=1,
		\end{align}
	\end{subequations}
	where $\mathbf{w}_{k,l}$ denotes $\mathbf{w}_k$ corresponding to the $l$-th group of positions. 
	Similarly, the optimal solution for $\hat{\bm{\rho}}_{k,l}$ is the eigenvector associated with the largest eigenvalue of matrix $\hat{\mathbf{X}}_{k,l}$, where
	\begin{align}\label{eq74}
		\hat{\mathbf{X}}_{k,l}&=\hat{\mathbf{B}}_{k,l}^{-\tfrac{1}{2}}\bm{\Gamma}_{2,l}^H\hat{\mathbf{T}}_{l}\mathbf{L}_{1,k}\hat{\mathbf{T}}_{l}\bm{\Gamma}_{2,l}\hat{\mathbf{B}}_{k,l}^{-\tfrac{1}{2}},\\\label{eq75}
		\hat{\mathbf{B}}_{k,l}&=\bm{\Gamma}_3^H\hat{\mathbf{T}}_{l}(\mathbf{L}_{4,k}+\mathbf{L}_{5,k})\hat{\mathbf{T}}_{l}\bm{\Gamma}_3.
	\end{align}
	Then, the set of all SSR values can be denoted as $\mathcal{S}=\{R_s^0,R_s^1,\ldots,R_s^{N_t-1}\}$, where $N_t = \lfloor\frac{N}{n_0}\rfloor$. Based on the block design principle, the top $\lfloor\frac{N_a}{n_0}\rfloor$ position groups with the largest SSR values are taken as candidate positions for MAs. However, the combined $\lfloor\frac{N_a}{n_0}\rfloor$ groups may violate the minimum distance constraint. Therefore, we retain only those groups satisfying the distance constraint and iteratively perform the same operation from the remaining candidate positions until $N_a$ valid MA positions are selected. For clarity, the detailed procedure is presented in Algorithm \ref{alg:alg2}. The complexity of Algorithm \ref{alg:alg2} is approximately $\tfrac{N_tn_0^3}{SN^3}$ that of the exhaustive search algorithm, where $S$ represents the maximum number of searches.
	
	\begin{algorithm}[t]
		\renewcommand{\algorithmicrequire}{\textbf{Input:}}
		\renewcommand{\algorithmicensure}{\textbf{Output:}}
		\caption{The proposed MA positions optimization method based on uniform grouping}\label{alg:alg2}
		\begin{algorithmic}[1]
			\REQUIRE $P_0$, $P_{\text{RIS}}$, $\mu_r$, $\mu_t$, $\sigma_k^2$, $\sigma_r^2$, $\sigma_e^2$, $N$, $N_a$, $M$, $K$, $\xi_1$, $\epsilon$.
			\ENSURE $\mathbf{T}$
			\STATE Initialize $\mathbf{u}_k$, $\mathbf{u}_{e,k}$, and $\boldsymbol{\Theta}$.
			\STATE Divide the $N$ candidate positions into $N_t$ groups.
			\STATE Optimize $\bm{\Theta}$ according to Algorithm \ref{alg:alg1} and compute SSR corresponding to $\hat{\mathbf{T}}_0$.
			\STATE \textbf{for} $n_t = 1:1 :N_t-1$ \textbf{do}
			\STATE\hspace{0.5cm} Calculate $\hat{\bm{\rho}}_{k,l}$ according to \eqref{eq73}. 
			\STATE\hspace{0.5cm} Update $\mathbf{w}_{k,l}=\sqrt{\tfrac{P_0}{K}}\bm{\Gamma}_{2,l}\hat{\bm{\rho}}_{k,l}$.
			\STATE\hspace{0.5cm} Calculate $R_s^{n_t}$ according to \eqref{eq20}.
			\STATE\hspace{0.5cm} \textbf{if} $n_t = N_t-1$ \textbf{then}
			\STATE\hspace{0.5cm}\hspace{0.5cm} Break.
			\STATE\hspace{0.5cm} \textbf{end if}
			\STATE\textbf{end for}
			\STATE Sort $\mathcal{S}=\{R_s^0,R_s^1,\ldots,R_s^{N_t-1}\}$ in descending order.
			\STATE\textbf{while} $\|\mathbf{T}\|_0\neq N_a$ \textbf{do}		
			\STATE\hspace{0.5cm} Select positions associated with the $\lfloor\frac{N_a}{K+1}\rfloor$ largest SSR values in $\mathcal{S}$ for MA placement.
			\STATE\hspace{0.5cm} Remove the set $\mathcal{S}^\prime$ of groups that do not satisfy the distance constraint.
			\STATE\hspace{0.5cm} Update $\|\mathbf{T}\|_0$.
			\STATE\hspace{0.5cm} Update $\mathcal{S}=\{\mathcal{S}|\mathcal{S}-\mathcal{S}^\prime\}$.
			\STATE\textbf{end while}
			\STATE\textbf{return} $\mathbf{T}$.
		\end{algorithmic}
	\end{algorithm}
	\subsubsection{Non-uniform grouping method}
	To explore optimal MA positioning, we propose a CS-based location optimization method through non-uniform position grouping. The initial state of $\mathbf{T}$ is set to $\mathbf{T}^0=\mathbf{I}_N$, from which the initial beamforming vectors can be obtained via Algorithm \ref{alg:alg1}. Leveraging key features extracted from the beamforming vectors, our method eliminates redundant candidate positions using CS techniques for group formation.
	
	Since all Bob share the same MA array, we introduce an auxiliary variable $\mathbf{W}$ to represent the set of all beamforming vectors, given by
	\begin{align}
		\begin{split}
			\mathbf{W}& = \mathbf{T}[\mathbf{w}_0,\mathbf{w}_1,\ldots,\mathbf{w}_{K-1}]\\
			&=\begin{bmatrix}t_0w_{_{0,0}}&t_0w_{_{1,0}}&\cdots&t_0w_{_{K-1,0}}\\
				t_1w_{_{0,1}}&t_1w_{_{1,1}}&\cdots&t_1w_{_{K-1,1}}\\
				\vdots&\vdots&\ddots&\vdots\\
				t_{_{N-1}}w_{_{0,N-1}}&t_{_{N-1}}w_{_{1,N-1}}&\cdots&t_{_{N-1}}w_{_{K-1,N-1}}
			\end{bmatrix},\\
			&=[\tilde{\mathbf{w}}_0^T,\tilde{\mathbf{w}}_1^T,\ldots,\tilde{\mathbf{w}}_{N-1}^T]^T,
		\end{split}
	\end{align}
	where $\tilde{\mathbf{w}}_n = t_n[w_{0,n},w_{1,n},\ldots,w_{K-1,n}]$ denote the weight vector corresponding to the $n$-th candidate position for $K$ Bob. For $\mathbf{T}^0=\mathbf{I}_N$, $\mathbf{W}^0 = [\mathbf{w}_0,\mathbf{w}_1,\ldots,\mathbf{w}_{N-1}]$ denotes $\mathbf{W}$ derived from the initial state. Now, define the $\ell_2$-norm set $\mathbf{w}=\big[\|\tilde{\mathbf{w}}_0\|^2,\|\tilde{\mathbf{w}}_1\|^2,\ldots,\|\tilde{\mathbf{w}}_{N-1}\|^2\big]^T$ of $\tilde{\mathbf{w}}_n$, and approximate the $\ell_0$-norm using $\ell_1$-norm.  According to \eqref{eq70}, we have
	\begin{align}
		\begin{split} \|\bm{\theta}^T\hat{\mathbf{H}}_k^{c}\cdot\sum_{i=0}^{K-1}|\mathbf{w}_i^T\mathbf{T}|\|^2&=\bm{\theta}^T\hat{\mathbf{H}}_k^{c}\text{diag}(\mathbf{w})(\bm{\theta}^T\hat{\mathbf{H}}_k^{c})^H
			\\&=\bm{\theta}^T\hat{\mathbf{H}}_k^{c}\text{diag}((\bm{\theta}^T\hat{\mathbf{H}}_k^{c})^H)\mathbf{w}.
		\end{split}
	\end{align}
	Then, the group sparsity problem based on CS can be described as
	\begin{subequations}\label{eq80}
		\begin{align}
			\text{P9:}\ &\underset{\mathbf{W}}{\min}\  \|\mathbf{w}\|_{1}\\
			&\mathrm{s.t.}\  \text{C20}: \bm{\theta}^T\hat{\mathbf{H}}_k^{c}\mathbf{W}=\bm{\theta}^T\hat{\mathbf{H}}_k^{c}\mathbf{W}^0,\forall k,\\
			&\quad\ \ \text{C21}:\bm{\theta}^T(\hat{\mathbf{H}}_e^c+\mathbf{h}_e^d)\mathbf{W}=\mathbf{0},
		\end{align}
	\end{subequations}
	where C20 and C21 represent the received signal patterns at all Bob and Eve, respectively, corresponding to the beamforming vectors of Bob $k$. 
	Through \eqref{eq80}, the positions corresponding to the non-zero elements in $\mathbf{w}$ are set as one group. Subsequently, the second group is generated from the positions corresponding to the zero-valued elements using CS, until all candidate positions are partitioned. The specific steps are illustrated as Algorithm \ref{alg:alg3}. The complexity of Algorithm \ref{alg:alg3} is approximately $\tfrac{\sum_{i=0}^{\hat{N}_t}{\hat{n}_i}^3}{SN^3}$ that of the exhaustive search algorithm, where $\hat{N}_t$ represents the number of groups and $n_i$ denotes the number of variables in the $i$-th CS.
	\begin{algorithm}[t]
		\renewcommand{\algorithmicrequire}{\textbf{Input:}}
		\renewcommand{\algorithmicensure}{\textbf{Output:}}
		\caption{The proposed MA positions optimization method based on non-uniform grouping}\label{alg:alg3}
		\begin{algorithmic}[1]
			\REQUIRE $P_0$, $P_{\text{RIS}}$, $\mu_r$, $\mu_t$, $\sigma_k^2$, $\sigma_r^2$, $\sigma_e^2$, $N$, $N_a$, $M$, $K$, $\xi_1$, $\epsilon$.
			\ENSURE $\mathbf{T}$
			\STATE Initialize $\mathbf{u}_k$, $\mathbf{u}_{e,k}$, and $\boldsymbol{\Theta}$.
			\STATE Set $\mathbf{T}^0=\mathbf{I}_N$, and then optimize $\mathbf{w}_k$ and $\bm{\Theta}$ using Algorithm \ref{alg:alg1}.
			\STATE Calculate $\bm{\theta}^T\hat{\mathbf{H}}_k^{c}\mathbf{W}^0$ and $\|\bm{\theta}^T\hat{\mathbf{H}}_k^{c}\cdot\sum_{i=0}^{K-1}|\mathbf{w}_i^T\mathbf{T}|\|^2$.		
			\STATE Initialize the number of iterations $n_t^\prime=0$.
			\STATE\textbf{while} $\|\mathbf{T}\|_0\neq N$ \textbf{do}
			\STATE\hspace{0.5cm} Set $n_t^\prime=n_t^\prime+1$.
			\STATE\hspace{0.5cm} Compute $\eqref{eq80}$ and set the positions corresponding to the non-zero values in $\mathbf{w}$ as one group.
			\STATE\hspace{0.5cm} Set the channels corresponding to the grouped positions to zero.
			\STATE\hspace{0.5cm} Normalize $\mathbf{w}_k = \tfrac{P_0\mathbf{w}_k}{K\|\mathbf{w}_k\|}$ and calculate $R_s^{n_t^\prime}$ according to \eqref{eq20}.
			\STATE\textbf{end while}	
			\STATE Sort $\hat{\mathcal{S}}=\{R_s^0,R_s^1,\ldots,R_s^{N_t^\prime-1}\}$ in descending order, where $N_t^\prime$ denotes the number of groups.
			\STATE\textbf{while} $\|\mathbf{T}\|_0\neq N_a$ \textbf{do}		
			\STATE\hspace{0.5cm} Select the corresponding position groups in sequence according to the SSR until $\|\mathbf{T}\|_0= N_a$.
			\STATE\hspace{0.5cm} Remove the set $\hat{\mathcal{S}}^\prime$ of groups that do not satisfy the distance constraint.
			\STATE\hspace{0.5cm} Update $\|\mathbf{T}\|_0$.
			\STATE\hspace{0.5cm} Update $\hat{\mathcal{S}}=\{\mathcal{S}|\hat{\mathcal{S}}-\hat{\mathcal{S}}^\prime\}$.
			\STATE\textbf{end while}
			\STATE\textbf{return} $\mathbf{T}$.
		\end{algorithmic}
	\end{algorithm}
	\subsection{Optimize $\alpha$, $\mathbf{u}_k$, and $\mathbf{u}_{e,k}$}
	With $\mathbf{w}_k$, $\alpha$ can be obtained. Based on the maximum ratio transmission (MRT), the beamforming vector corresponding to Bob $k$ can be expressed as
	\begin{equation}\label{eq81}
		\mathbf{v}_{k}^{\text{MRT}}=\tfrac{\mathbf{T}\mathbf{Q}_k^H\mathbf{u}_k}{\|\mathbf{T}\mathbf{Q}_k^H\mathbf{u}_k\|}.
	\end{equation}
	Under MRT, the desired signal received by Bob $k$ can be represented by $y_k^{\text{MRT}}=N_as_k$. However, it is difficult to achieve full array gain due to practical limitations such as multi-user interference, HWIs, and imperfect CSI. Using the obtained \( \mathbf{w}_k \), the useful signal received at Bob $k$ can be expressed as
	\begin{equation}\label{eq82}
		\begin{split}
			y_k^{\text{MA}}&=\mathbf{u}_k^H\mathbf{Q}_k\mathbf{T}\mathbf{w}_ks_k\\&=\frac{\|\mathbf{u}_k^H\mathbf{Q}_k\mathbf{T}\mathbf{w}_k\|}{N_a}y_k^{\text{MRT}}e^{j\mathbf{u}_k^H\mathbf{Q}_k\mathbf{T}\mathbf{w}_k}.
		\end{split}
	\end{equation}
	According to \eqref{eq38} and \eqref{eq82}, the beamforming vector $\mathbf{v}_k$ corresponding to Bob $k$ can be denoted as
	\begin{equation}\label{eq83}
		\begin{split}
			\mathbf{v}_k = \sqrt{\frac{K}{\alpha P_0}}\frac{\|\mathbf{u}_k^H\mathbf{Q}_k\mathbf{T}\mathbf{w}_k\|}{N_a\|\mathbf{T}\mathbf{Q}_k^H\mathbf{u}_k\|}\mathbf{T}\mathbf{Q}_k^H\mathbf{u}_ke^{j\mathbf{u}_k^H\mathbf{Q}_k\mathbf{T}\mathbf{w}_k}.
		\end{split}
	\end{equation}
	Using $\mathbf{v}_k^H\mathbf{T}\mathbf{v}_k = 1$, the value of \( \alpha \) can be calculated as
	\begin{equation}\label{eq84}
		\begin{split}
			\mathbf{\alpha}=\frac{K\|\mathbf{u}_k^H\mathbf{Q}_k\mathbf{T}\mathbf{w}_k\|^2}{P_0N_a^2\|\mathbf{T}\mathbf{Q}_k^H\mathbf{u}_k\|^2}.
		\end{split}
	\end{equation}
	According to \eqref{eq84}, the value of $\alpha$ is related to the number of Bob, the received signal power, the transmit power, the number of MAs, and the channel conditions of Bob. Given the transmit power, the number of mobile antennas, and the channel conditions of Bob, increasing the number of Bob and enhancing the received power can increase the energy allocated to CM transmission. A larger value of $\alpha$ is conducive to achieving better robustness, as analyzed in \eqref{eq43}. Substituting \eqref{eq83} into ${\mathbf{w}}_k=\sqrt{\tfrac{\alpha P_0}{K}}\mathbf{v}_k+\sqrt{\tfrac{(1-\alpha) P_0}{K}}\mathbf{v}_{a,k}zs_k^*$, we have
	\begin{align}
		\begin{split}
			\mathbf{v}_{a,k}= &\sqrt{\tfrac{K}{(1-\alpha) P_0}}\\&(\mathbf{w}_ks_kz^*-\tfrac{\|\mathbf{u}_k^H\mathbf{Q}_k\mathbf{T}\mathbf{w}_k\|}{N_a\|\mathbf{T}\mathbf{Q}_k^H\mathbf{u}_k\|}\mathbf{T}\mathbf{Q}_k^H\mathbf{u}_ke^{j\mathbf{u}_k^H\mathbf{Q}_k\mathbf{T}\mathbf{w}_k}s_kz^*).
		\end{split}
	\end{align}
	
	Given $\mathbf{w}_k$, $\boldsymbol{\Theta}$, and $\mathbf{T}$, $\mathbf{u}_k$ and $\mathbf{u}_{e,k}$ can be calculated according to minimum mean square error (MMSE) theorem. The MSE matrix of estimation corresponding to Bob $k$ can be expressed as
	\begin{align}
		\mathbf{E}_k(\mathbf{u}_k) &= \mathbb{E}\{(\hat{y}_k-s_k)(\hat{y}_k-s_k)^H\}\notag\\
		&=\tfrac{{\alpha P_0}}{K}\sum_{i=0}^{K-1}(1+\mu_{r})\mathbf{u}_k^H\mathbf{Q}_k\mathbf{T}\mathbf{v}_i\mathbf{v}_i^H\mathbf{T}^H\mathbf{Q}_k^H\mathbf{u}_k+\notag\\
		&\quad\tfrac{{(1-\alpha) P_0}}{K}\sum_{i=0}^{K-1}(1+\mu_{r})\mathbf{u}_k^H\mathbf{Q}_k\mathbf{T}\mathbf{v}_{a,i}\mathbf{v}_{a,i}^H\mathbf{T}^H\mathbf{Q}_k^H\mathbf{u}_k+\notag\\
		&\quad(1+\mu_{r})\mu_{t}\mathbf{u}_{k}^H\mathbf{Q}_{k}\mathbf{T}\mathbf{R}_t\mathbf{T}^H\mathbf{Q}_{k}^H\mathbf{u}_{k}+\notag\\
		&\quad\sigma_r^{2}\mathbf{u}_k^H\mathbf{F}_{k}^H\boldsymbol{\Theta}\boldsymbol{\Theta}^H\mathbf{F}_{k}\mathbf{u}_k+(1+\mu_{r})\sigma_k^{2}\mathbf{u}_k^H\mathbf{u}_k-\notag\\
		&\quad\sqrt{\tfrac{\alpha P_0}{K}}(\mathbf{u}_{k}^H\mathbf{Q}_k\mathbf{T}\mathbf{v}_k+\mathbf{v}_k^H\mathbf{T}\mathbf{Q}_k^H\mathbf{u}_{k})+1\notag\\
		&=\mathbf{u}_k^H\mathbf{E}_1\mathbf{u}_k-\sqrt{\tfrac{\alpha P_0}{K}}\mathbf{v}_k^H\mathbf{T}\mathbf{Q}_k^H\mathbf{u}_{k}+E_2,
	\end{align}
	where 
	\begin{align}
		\mathbf{E}_1 &= \tfrac{{\alpha P_0}}{K}\sum_{i=0}^{K-1}(1+\mu_{r})\mathbf{Q}_k\mathbf{T}\mathbf{v}_i\mathbf{v}_i^H\mathbf{T}^H\mathbf{Q}_k^H+\notag\\
		&\quad\tfrac{{(1-\alpha) P_0}}{K}\sum_{i=0}^{K-1}(1+\mu_{r})\mathbf{Q}_k\mathbf{T}\mathbf{v}_{a,i}\mathbf{v}_{a,i}^H\mathbf{T}^H\mathbf{Q}_k^H+\notag\\
		&\quad(1+\mu_{r})\mu_{t}\mathbf{Q}_{k}\mathbf{T}\mathbf{R}_t\mathbf{T}^H\mathbf{Q}_{k}^H+\notag\\
		&\quad\sigma_r^{2}\mathbf{F}_{k}^H\boldsymbol{\Theta}\boldsymbol{\Theta}^H\mathbf{F}_{k}+(1+\mu_{r})\sigma_k^{2}\mathbf{I}_{N_k},\\
		E_2 &= -\sqrt{\tfrac{\alpha P_0}{K}}\mathbf{u}_{k}^H\mathbf{Q}_k\mathbf{T}\mathbf{v}_k+1.
	\end{align}
	The derivative of $\mathbf{E}_k(\mathbf{u}_k)$ w.r.t. $\mathbf{u}_k$ equals
	\begin{align}
		\frac{\partial \mathbf{E}_k(\mathbf{u}_k)}{\partial\mathbf{u}_k}=	\mathbf{E}_1^T\mathbf{u}_k^*-\sqrt{\tfrac{\alpha P_0}{K}}\mathbf{Q}_k^*\mathbf{T}\mathbf{v}_k^*.
	\end{align}
	Let $\frac{\partial \mathbf{E}_k(\mathbf{u}_k)}{\partial\mathbf{u}_k}=0$, and $\mathbf{u}_k$ can be calculated as
	\begin{align}
		{\mathbf{u}_k}=	\sqrt{\tfrac{\alpha P_0}{K}}\mathbf{E}_1^{-H}\mathbf{Q}_k\mathbf{T}\mathbf{v}_k.
	\end{align}
	Similarly, the receive beamforming vector $\mathbf{u}_{e,k}$ corresponding to Eve can be denoted as
	\begin{align}
		{\mathbf{u}_{e,k}}=	\sqrt{\tfrac{\alpha P_0}{K}}\mathbf{E}_3^{-H}\mathbf{Q}_e\mathbf{T}\mathbf{v}_k,
	\end{align}
	where
	\begin{align}
		\mathbf{E}_3 &= \tfrac{{\alpha P_0}}{K}\sum_{i=0}^{K-1}\mathbf{Q}_e\mathbf{T}\mathbf{v}_i\mathbf{v}_i^H\mathbf{T}^H\mathbf{Q}_e^H+\notag\\
		&\quad\tfrac{{(1-\alpha) P_0}}{K}\sum_{i=0}^{K-1}\mathbf{Q}_e\mathbf{T}\mathbf{v}_{a,i}\mathbf{v}_{a,i}^H\mathbf{T}^H\mathbf{Q}_e^H+\notag\\
		&\quad\mu_{t}\mathbf{Q}_{k}\mathbf{T}\mathbf{R}_t\mathbf{T}^H\mathbf{Q}_{k}^H+\sigma_r^{2}\mathbf{F}_{e}^H\boldsymbol{\Theta}\boldsymbol{\Theta}^H\mathbf{F}_{e}+\sigma_e^{2}\mathbf{I}_{N_e}.
	\end{align}
	\section{Simulation Results}\label{sec:5}
	In this section, simulation results are presented to validate the effectiveness of the proposed method. Simulation parameters are set as follows: Bob and Eve each carry $N_k=N_e=4$ linear antennas with fixed positions, and the minimum spacing of the MAs carried by the BS is $d = \lambda/2$. The system operates at frequency $f = 15\ \text{GHz}$, serving $K=2$ Bob located at $[0,5,0]$ and $[0,15,0]$, while an Eve is detected at $[2, 30, 0]$. The coordinates are in meters. The BS and RIS are located at $[0,0,2]$ and $[-1,10,1]$, respectively. The aperture of the RIS is set to be greater than 0.23 meters to comply with the NF model. The noises at Bob, Eve, and RIS are set as $\sigma_k^2=-60$~dBm, $\sigma_e^2=-60$~dBm, and $\sigma_r^2=-70$~dBm, respectively. The upper bounds of the channel estimation errors corresponding to Bob $k$ and Eve are $\varepsilon_k=0.01$ and $\varepsilon_e=0.02$, respectively. Unless otherwise specified, the transmission power, the number of RIS units, reflection power threshold, and the HWIs parameters are set to $P_0 = 30$ dBm, $M=625$,  $P_{\text{RIS}} = 10$ dBm, and $\mu_t=\mu_r=0.01$, respectively.
	\begin{figure}[t]
		\centering
		\includegraphics[width=0.45\textwidth, trim = 2 1 1 10,clip]{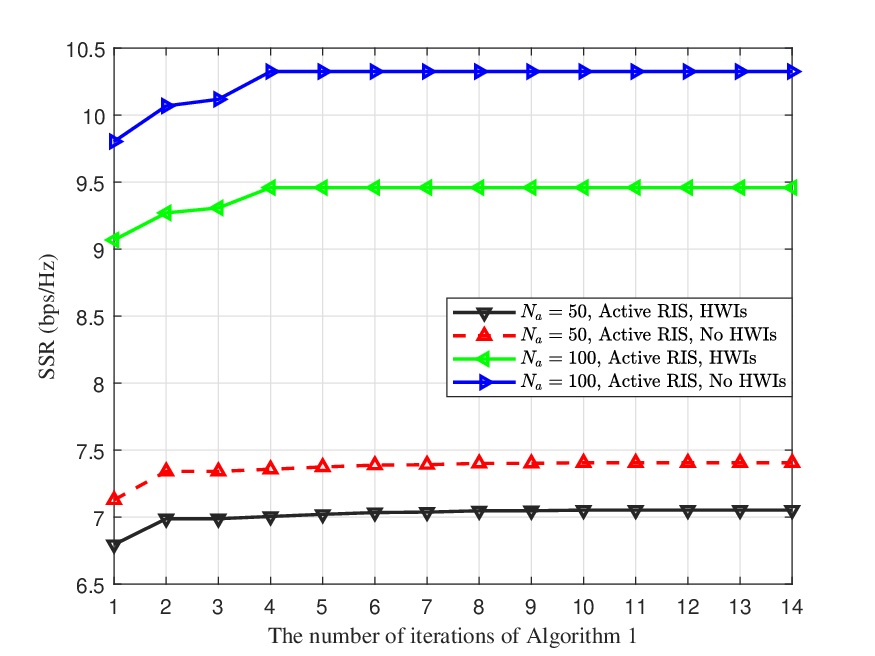}\\
		\caption{Convergence behaviour of the Algorithm \ref{alg:alg1}.}\label{fig3}
	\end{figure}
	
	Firstly, the convergence behaviour of the Algorithm \ref{alg:alg1} is illustrated in Fig. \ref{fig3}. It can be observed that Algorithm \ref{alg:alg1} is capable of converging under various conditions, including the presence or absence of HWIs, as well as different numbers of MAs. In practice,  the achievable SSR performance is degraded due to the existence of HWIs.
	
	\begin{figure}[t]
		\centering
		\includegraphics[width=0.45\textwidth, trim = 2 1 1 10,clip]{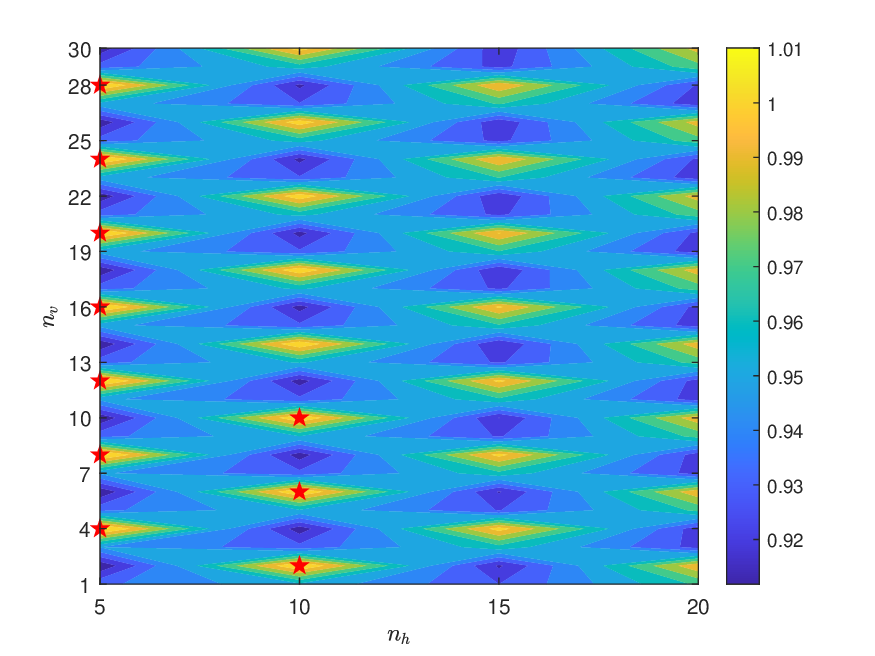}\\
		\caption{The SSR distribution across all groups obtained via Algorithm \ref{alg:alg2}.}\label{fig4}
	\end{figure}
	\begin{figure}[t]
		\centering
		\includegraphics[width=0.45\textwidth, trim = 2 1 1 10,clip]{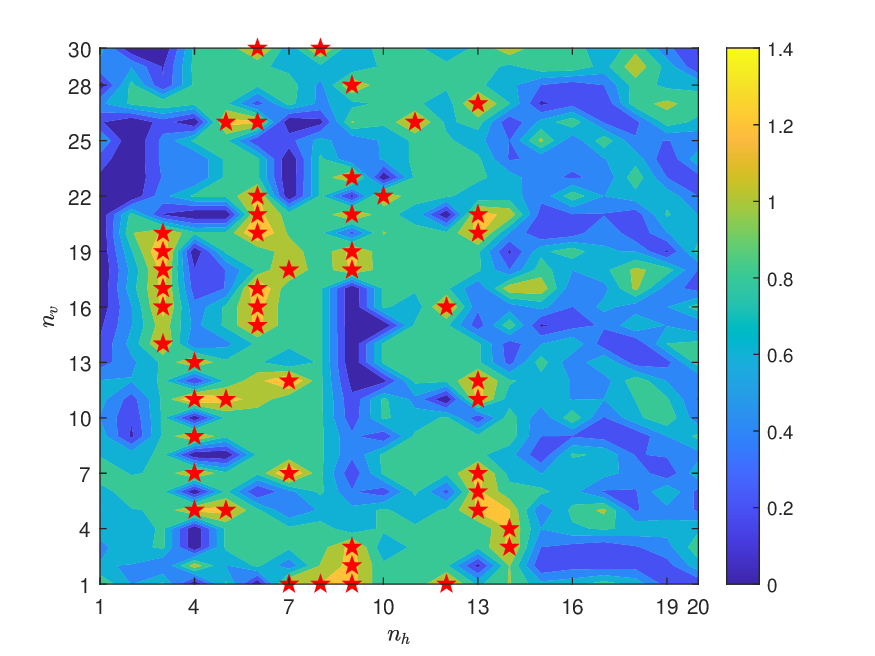}\\
		\caption{The SSR distribution of all candidate positions obtained via Algorithm \ref{alg:alg3}.}\label{fig5}
	\end{figure}
	Figs. \ref{fig4} and \ref{fig5} show the SSR distributions obtained using Algorithm \ref{alg:alg2} and Algorithm \ref{alg:alg3}, respectively. Red pentagram markers in Fig. \ref{fig4} indicate group locations, while in Fig. \ref{fig5} they represent MA positions. In this configuration, the MR is set to $\mathcal{C}_B = 0.19\ m\times0.29\ m$, with 600 candidate positions and 50 MAs being considered. Under uniform grouping, groups with high SSR values exhibit uniform distribution and alternate with groups corresponding to low SSR values. Compared with uniform grouping, non-uniform grouping achieves superior SSR performance with non-uniformly distributed optimal positions for MAs. Fig. \ref{fig5} reveals that certain candidate locations exhibit significantly low SSR values, indicating that they are redundant. The observed SSR degradation stems from multiple factors including severe HWIs at Bob, strong multi-user interference, elevated RIS reflection noise, and AN power substantially exceeding the information-bearing signal power. Furthermore, the uniform grouping array in Fig. \ref{fig4} can be conceptually decomposed into multiple small FPA subarrays, and the non-uniform grouping array in Fig. \ref{fig5} comprises multiple small MA subarrays. This structural perspective demonstrates that non-uniform MA arrangements within constrained MRs deliver enhanced SSR performance.
	
	\begin{figure}[t]
		\centering
		\includegraphics[width=0.45\textwidth, trim = 2 1 1 10,clip]{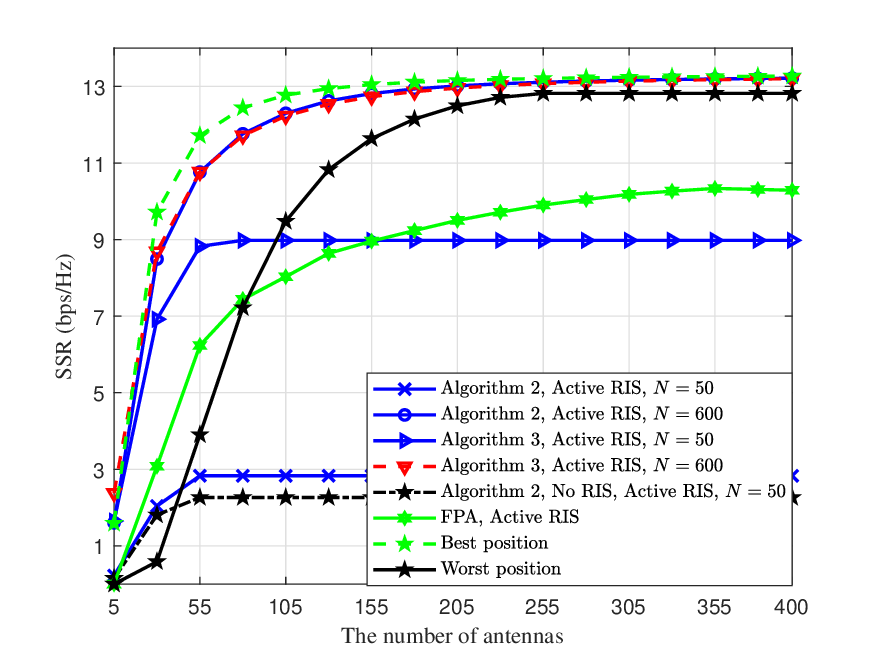}\\
		\caption{SSR versus the number of antennas.}\label{fig6}
	\end{figure}
	Figs. \ref{fig4} and \ref{fig5} reveal that, in addition to the selected candidate positions, there exist suboptimal SSR positions that remain unselected. To comprehensively investigate the performance gains offered by the proposed algorithms, Fig. \ref{fig6} illustrates the SSR performance versus the MA quantity. Four benchmark schemes are established: (1) ``\textbf{No RIS}'', (2) ``\textbf{FPA}'', (3) ``\textbf{Best position}'', and (4) ``\textbf{Worst position}''. The ``\textbf{Best position}'' and ``\textbf{Worst position}'' represent the optimal and poorest SSR performances, respectively, obtained from 10000 random position selections. Notably, we employ the ``\textbf{Best position}'' configuration as a benchmark alternative to exhaustive search, given the prohibitive computational complexity of the latter. For instance, selecting 50 MA positions from 600 candidates would require $\tfrac{600!}{50!(600-50)!}$ search operations. Fig. \ref{fig6} demonstrates that SSR performance for both benchmark and proposed schemes improves with increasing antenna count until reaching a plateau. Algorithm \ref{alg:alg3} outperforms Algorithm \ref{alg:alg2} when the number of candidate positions is limited, while achieving comparable performance given sufficient candidate positions. Notably, Algorithm \ref{alg:alg3} maintains superior performance over Algorithm \ref{alg:alg2} with fewer MAs. The proposed algorithms consistently surpass the ``\textbf{Worst position}'' benchmark, and approach ``\textbf{Best position}'' performance when deploying 205 MAs. Compared to ``\textbf{FPA}'', the proposed algorithms can achieve 28\% SSR enhancement while requiring approximately 150 fewer antennas. Moreover, in MA systems, active RIS deployment demonstrates superior SSR performance compared to ``\textbf{No RIS}''. These results conclusively demonstrate that MA arrays can significantly enhance system performance, delivering superior security with reduced antenna count.
	\begin{figure}[t]
		\centering
		\includegraphics[width=0.45\textwidth, trim = 2 1 1 10,clip]{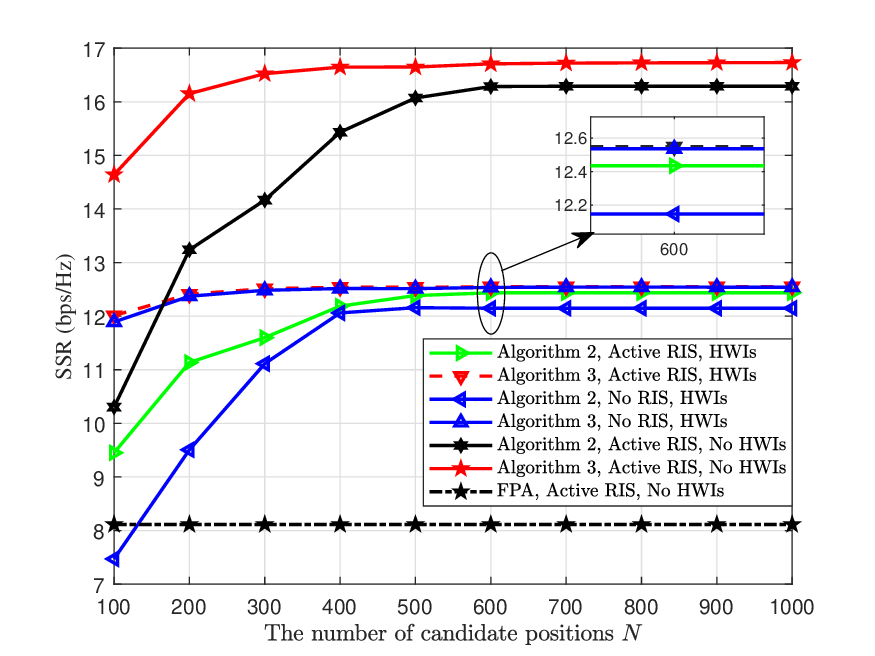}\\
		\caption{SSR versus the number of candidate positions.}\label{fig7}
	\end{figure}
	\begin{figure}[t]
		\centering
		\includegraphics[width=0.45\textwidth, trim = 2 1 1 10,clip]{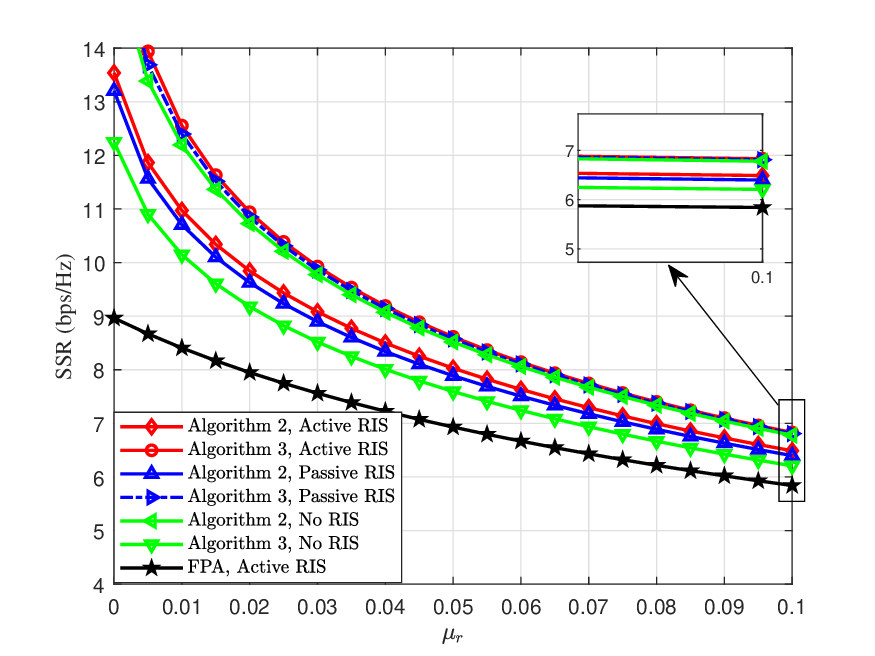}\\
		\caption{SSR versus the HWI parameter $\mu_r$.}\label{fig8}
	\end{figure}
	Fig. \ref{fig7} illustrates the SSR performance versus the number of candidate positions. The results demonstrate that SSR performance improves with increasing candidate positions, reaching saturation when the number exceeds 600. This stabilization occurs because additional candidate positions provide diminishing returns in DoF ($\geq0$). Notably, compared to Algorithm \ref{alg:alg2}, Algorithm \ref{alg:alg3} shows limited performance gain over ``\textbf{No RIS}'', primarily due to increased HWIs and reflection noise power caused by high RIS reflection power. Based on this, a fifth benchmark scheme is added, i.e., ``\textbf{No HWIs}''. Under conditions of high HWIs (where $\mu_r=\mu_t=0.01$), the proposed Algorithm \ref{alg:alg3} exhibits a maximum performance degradation of 24\% compared to ``\textbf{No HWIs}''. Further analysis in Fig. \ref{fig8} examines SSR performance under different values of parameter $\mu_r$ (where $\mu_r = \mu_t$). The results indicate that lower values of $\mu_r$ enhance security performance. From a practical implementation perspective, developing cost-effective transceivers with low HWI parameters would be beneficial.
	
	\begin{figure}[t]
		\centering
		\includegraphics[width=0.45\textwidth, trim = 2 1 1 10,clip]{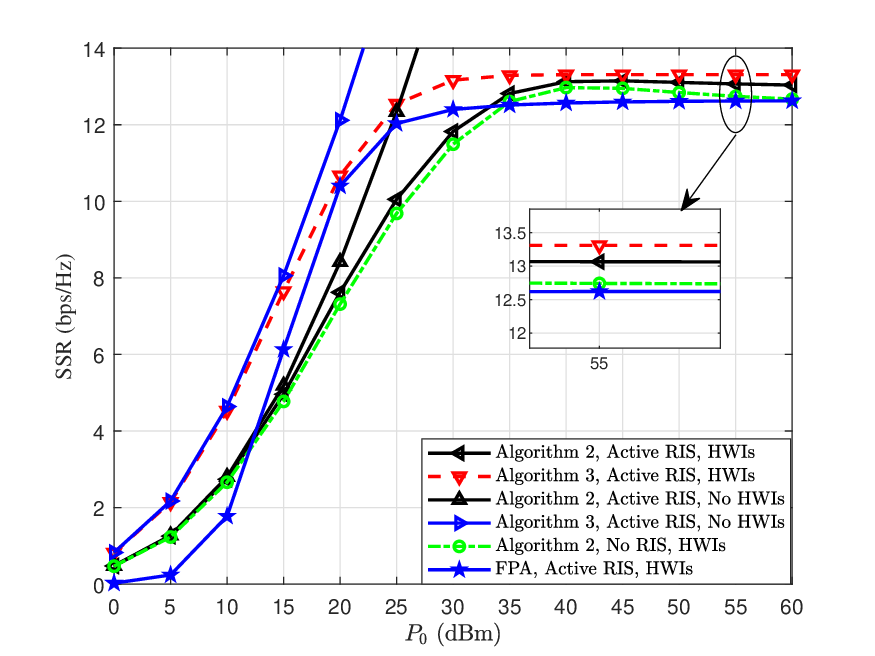}\\
		\caption{SSR versus the transmission power $P_0$.}\label{fig9}
	\end{figure}
	
\begin{figure}[t]
		\centering
		\includegraphics[width=0.45\textwidth, trim = 2 1 1 10,clip]{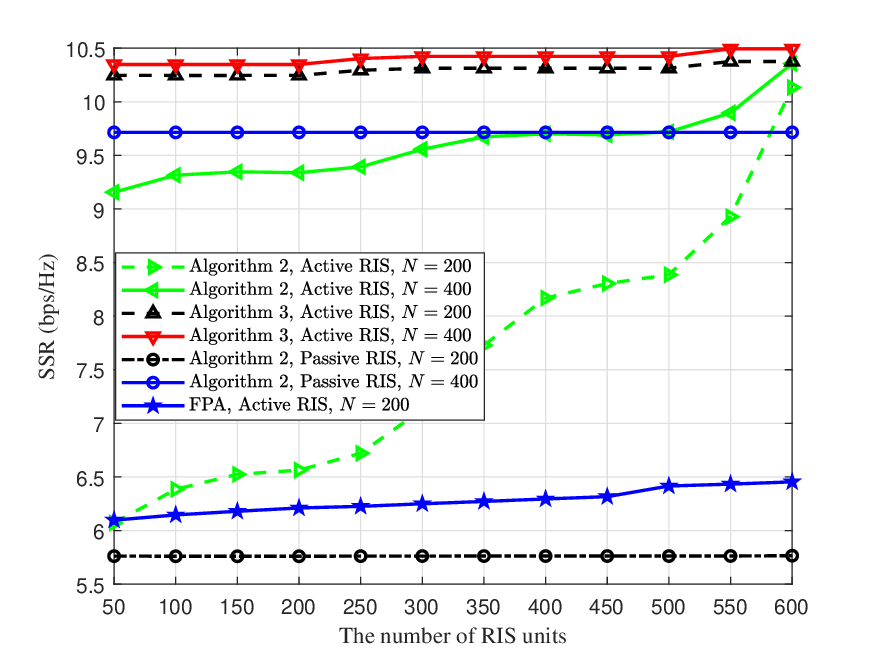}\\
		\caption{SSR versus the number of RIS units.}\label{fig10}
	\end{figure}

	Figs \ref{fig9} and \ref{fig10} respectively illustrate the SSR performance versus transmit power and RIS unit count. As the transmit power increases, the SSR performance under HWIs gradually stabilizes. The proposed Algorithm \ref{alg:alg3} consistently achieves superior security performance compared to ``\textbf{FPA}''. However, the SSR growth rate obtained by Algorithm \ref{alg:alg2} between 10 dBm and 20 dBm is slower than that of ``\textbf{FPA}'', resulting in inferior performance within this range, while maintaining advantages below 12 dBm and above 34 dBm. With increasing RIS units, both proposed algorithms show improved SSR performance across different candidate position quantities. Notably, Algorithm \ref{alg:alg3} exhibits limited performance gains with additional RIS elements. Consequently, we introduce a sixth benchmark ``\textbf{Passive RIS}'' for comparison. In passive RIS-assisted systems, the SSR performance remains nearly constant regardless of RIS unit count, as the severe multiplicative fading in RIS reflection paths causes the transmitter to concentrate power primarily on the direct path.
	
\begin{figure}[t]
		\centering
		\includegraphics[width=0.45\textwidth, trim = 2 1 1 10,clip]{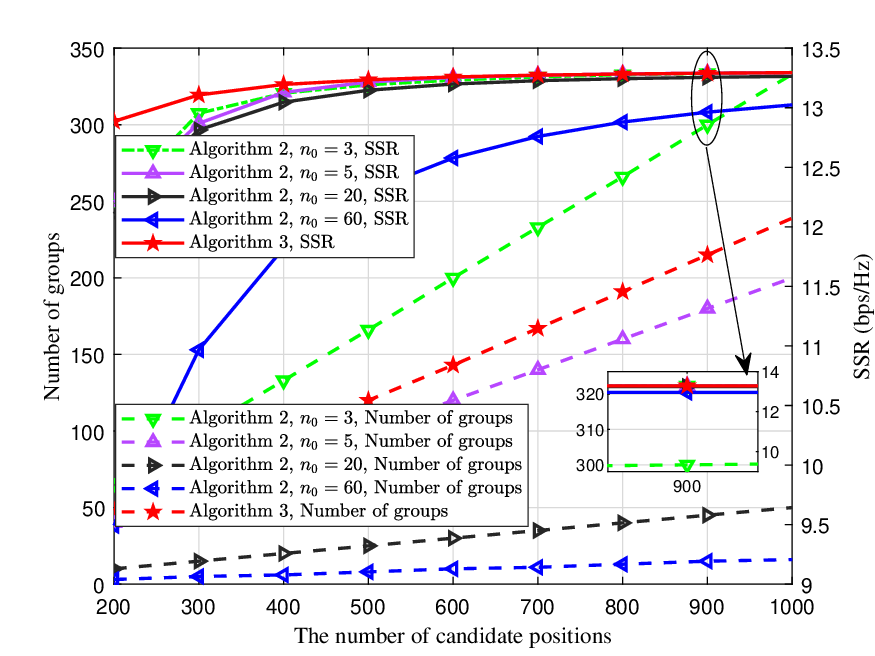}\\
		\caption{SSR and the number of groups versus the number of candidate positions.}\label{fig11}
	\end{figure}

	Fig. \ref{fig11} provides a comparative performance analysis between the proposed Algorithm \ref{alg:alg2} and Algorithm \ref{alg:alg3}. For a fixed total number of candidate positions, increasing the per-group candidate positions in Algorithm \ref{alg:alg2} leads to reduced group quantity and consequently degraded SSR performance. When $n_0=3$, although Algorithm \ref{alg:alg2} maintains a larger number of groups than Algorithm \ref{alg:alg3}, its performance remains inferior.
	\section{Conclusions}\label{sec:6}
	This paper presented a novel near-field directional modulation design for RIS-assisted MA systems. We addressed the challenging non-convex SSR maximization problem through joint optimization of transmit beamforming vectors, phase shift matrices, power allocation factors, discrete MA positions, and receive beamforming vectors. To solve this problem, we develop three low-complexity algorithms: an iterative beamforming and phase shift optimization method based on leakage theory and phase alignment, along with two MA positioning algorithms employing uniform grouping and CS-based non-uniform grouping strategies. Simulation results demonstrate that our proposed algorithms achieve 28\% higher SSR performance while requiring 37.5\% fewer antenna elements compared to conventional fixed-position antenna systems, validating their effectiveness.

%
%
%
%
%
%
%

\ifCLASSOPTIONcaptionsoff
  \newpage
\fi

\renewcommand\refname{References}
\bibliographystyle{IEEEtran}
\bibliography{mybib}

%
%
%

\end{document}